\documentclass[preprint, authoryear,3p, times, 12pt]{elsarticle}
\usepackage[utf8]{inputenc}
\usepackage{amsmath}
\usepackage{amssymb}
\usepackage{epsfig}
\usepackage{url}
\usepackage{subfigure}
\usepackage{multirow}
\usepackage{rotating}
\usepackage{booktabs}
\usepackage{algorithm}
\usepackage{algpseudocode}

\begin{document}

\title{KPCA Spatio-temporal trajectory point cloud classifier for recognizing human actions in a CBVR system}

\author{Iván Gómez-Conde${^1}$, David N. Olivieri${^1}$ \\
\small $^{1}$Department of Computer Science, University of Vigo, Ourense 32004, Spain\\
ivangconde@uvigo.es, olivieri@ei.uvigo.es
}

\begin{frontmatter}
\begin{abstract}
We describe a content based video retrieval (CBVR) software system for identifying specific locations of a human action within a full length film, and retrieving similar video shots from a query.   For this, we introduce the concept of a trajectory point cloud for classifying unique actions, encoded in a spatio-temporal covariant eigenspace,  where  each point is characterized by its spatial location, local Frenet-Serret vector basis, time averaged curvature and torsion and the mean osculating hyperplane.  Since each action can be distinguished by their unique trajectories within this space, the trajectory point cloud is used to define an adaptive distance metric for classifying queries against stored actions.  Depending upon the distance to other trajectories, the distance metric uses either large scale structure of the trajectory point cloud, such as the mean distance between cloud centroids or the difference in hyperplane orientation, or small structure such as the time averaged curvature and torsion,  to classify individual points in a fuzzy-KNN.  Our system can function in real-time and has an accuracy greater than  93\%  for multiple action recognition within video repositories. We demonstrate the use of our CBVR system in two situations:  by locating specific frame positions of trained actions in two full featured films, and video shot retrieval from a database with a web search application. 
\end{abstract}
\end{frontmatter}

\section{Introduction}

Recognizing specific human activities from real-time or recorded videos is a challenging practical problem in computer vision research.  If implemented as an efficient search engine, such algorithms would be valuable for managing and querying large multimedia database repositories where keyword searches on human actions are practically meaningless.  Thus, a \emph{content-based video retrieval} (CBVR) paradigm, where such queries are undertaken by comparing the actual multimedia content - as distinguished from others that compare only semantic tags - can provide a more powerful indexing/annotation and retrieval methods to produce richer query results.  Within this paradigm, computer vision algorithms are used to automatically index videos along the entire timeline consisting of semantics and feature vectors. Queries compare a similar reduction of the input video to all those feature vectors in the video repository.  To be practically useful as a search engine, the CBVR algorithms must be fast, robust, and accurate. 

We describe a CBVR system and algorithms for recognizing human actions in stored or real-time streaming videos by using a velocity encoded spatio-temporal representation of the movement.  In our method, each image frame of the original video shot is replaced by a simplified image, called an MVFI (motion vector flow instances) motion template, that extracts the direction and strength of the velocity flow field  \citep{Olivieri2012},  found by frame-to-frame differencing. Each template image can be further projected as a point into a reduced dimensional eigenspace through a Principal component analysis (PCA), or kernel-PCA (KPCA)  transformation,  so that frames of the entire video sequence, when projected into this space, trace out a unique curve we call the \emph{spatio-temporal trajectory}.  These trajectories provide an efficient technique for distinguishing different actions since similar actions have similar trajectories, while different actions can have radically different trajectories.  

For comparing trajectories, we describe a novel classification scheme that uses local differential geometric properties of these curves. We refer to our algorithm as the \emph{trajectory point cloud} classifier method.  In this method,  each n-dimensional point  contains information about the local neighborhood of the trajectory, while the collection of such points defines a macroscopic object, or cloud. In this way, the algorithm works on two scales for determining the distance between different trajectories (or actions). For large trajectory separations,  the distance is dominated by the difference of centroids between clouds.  For partially overlapping trajectories, the distance is dominated by a mean hyperplane that defines unique orientations of the trajectories, and when there is significant overlap, difference between the  orientation of local patches of the trajectory dominate the distance metric.  This description is valid since different types of actions will lie on completely different osculating hyperplanes, while similar actions would lie on the same plane, or one that is close-by.  At a small scale, individual points in the trajectory point cloud are endowed with their local geometric properties of their part of the curve in which they are embedded.  This local information can be used to infer class membership.   We show that our distance metric is more effective than traditional classifiers based upon methods such as KNN that do not incorporate information about the connectedness of points.   Moreover, with this technique, we remediate one of the traditional drawbacks of spatio-temporal methods - that they are limited to describing global properties of the motion. By utilizing this local information, slight differences of the movement can be distinguished.

\begin{figure*}[htb]
\begin{center}  
\epsfig{file=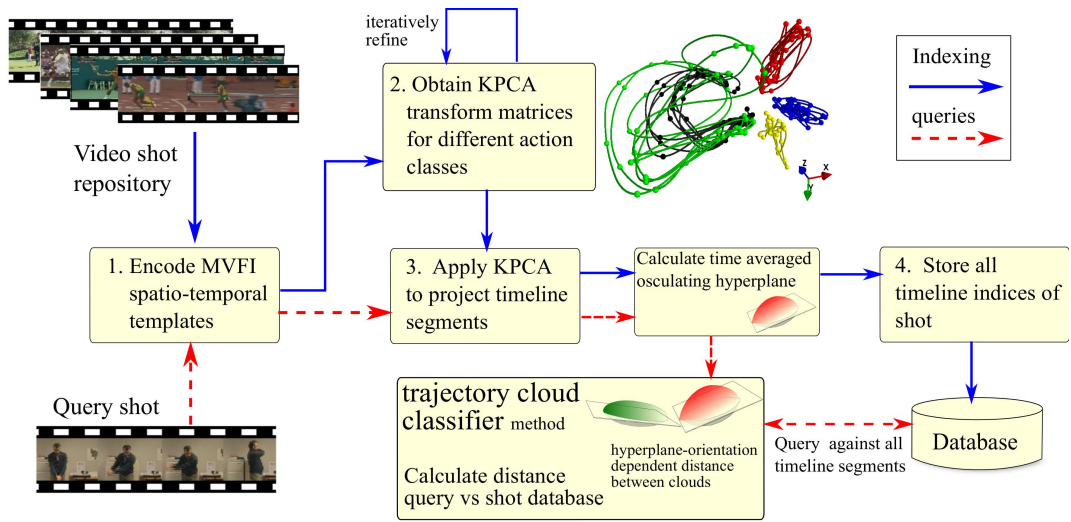, width=0.85\textwidth, angle=0}
\caption{\emph{Trajectory point cloud classifier}: Major steps of our CBVR system for storing and retrieval of  human actions from a set of videos. The \emph{indexing} phase produces velocity encoded KPCA spatio-temporal trajectories for discrete points along the shot timeline.  Our classification method also determines the \emph{time averaged osculating hyperplane} (see text for description).  This data is stored in the database.  The \emph{querying} phase performs the same processing steps, but additionally compares the hyperplanes and spatio-temporal feature vectors to all the shots in the database in order to produce similarity scores.}
\label{fig:workflow}   
\end{center}
\end{figure*}

Figure \ref{fig:workflow} shows a block diagram of our CBVR recognition system, consisting of two separate, but interconnected branches:  the \emph{indexing} and the \emph{querying} path.   In the indexing step, a set of videos are processed using computer vision algorithms with the purpose of annotating human actions (e.g., walking, jogging, jumping, etc.) along the timeline of a video.  In our implementation, indexing consists of obtaining KPCA spatio-temporal trajectories from an encoding of the velocity field at evenly distributed points along the timeline of the video, from a moving window of overlapping video segments. As shall be described in this paper, because spatio-temporal points are connected, the trajectory point cloud allows us to obtain information about the mean hyperplane on which this trajectory lives.  This meta- information associated with a particular video shot is stored in the database.  In the \emph{querying} phase,  an input video is processed with the same steps as the indexed videos and comparisons are made between all metadata of the query with all metadata from all videos in the repository. 

This paper is organized as follows.  First, we briefly review previous work on human action recognition.  Relevant mathematical details of constructing the linear-PCA and kernel-PCA covariance eigenspaces are provided and comparisons are carried out with two human action databases,  the KTH \citep{Schuldt2004} and the  MILE \citep{Olivieri2012}.  Next, we describe our new \emph{trajectory cloud} classifier method that is capable of resolving ambiguities that can arise for recognizing different types of action classes. Finally, we describe two examples of our CBVR system: locating specific positions along the timeline of a set of full feature films that contain particular human actions, and as a search engine for retrieving similar videos from a video shot repository.

\section{Background}
\label{sec:sec2}

Characterizing human motion without markers is a difficult computer vision problem that has generated a large amount of research.  Many different approaches in this field cover a broad spectrum of techniques - ranging from full-body tracking in 3D with multiple cameras \citep{Rius2009} to Bayesian inference models \citep{Meeds2008}.  A recent review \citep{Poppe10} and new textbook \citep{Szeliski2010} provide a taxonomic overview of the most salient algorithms that have been developed to characterize human motion. Methods also vary greatly in computational requirements, so that a solution which is more precise may be practically unusable for real-time applications or for a search engine that will be used in a CBVR system  \citep{Hu2011, Hosseini2013}.

Several recent reviews of content based video indexing and retrieval are available \citep{ Beecks2010, Bhatt2011, Hu2011}.  Specific CBVRs for retrieving video shots from a query with human actions have been described by \citet{Jones2013} and in \citep{Laptev2008} by using a full movie repository.  A large scale data mining methods that uses unsupervised clustering of human action videos was described by \citep{Liao2013}.  Another large scale study, that could treat more than 100 human actions, as been reported by \citet{Nga2014}.  This study automatically extracting video shots from semantic queries by using videos examples that have been previously tagged.  Many other specific studies exist in more narrow domains, such as that by \citet{Kucuk2011}, who described a system for indexing and querying news videos. Nonetheless, all CBVR methods to date grapple with the spatio-temporal problem, that does not exist in content based image retrieval.  Another common theme in all studies is the necessity of CBVR systems to treat the large amount variations of actions in videos.   For this reason, universally applicable CBVR systems are still in their infancy.

\subsection{Computationally intensive methods}

Amongst the most computationally demanding solutions are those that capture the full human body part motion over time, such as work described in \citep{Rius2009}, where the 3D tracking was accomplished with particle filters or \citep{Sadek2013}, where affine invariant features are derived from 3D spatio-temporal action shapes. Another example is work by \citep{Ugolotti2013}, where a particle swarm model is used for detecting people and performing 3D reconstruction. In another approach, full body motion is inferred from a  probabilistic graphical model that determines connected sticks figures \citep{Meeds2008}.  Similarly,  \citep{Felzenszwalb2010} describe a multiscale deformable parts model based upon segmenting human parts from each image frame.   While many of these methods are able to capture fine details of body motion,  they would require excessive computation, rendering them unusable for real-time information retrieval.  Also, the low level information requires another level of processing to distinguish actions.  One example where this low-level parts movement information is converted into higher level information is provided in \citep{Ikizler2008}, who used Hidden Markov models (HMMs) to infer composite human motion/actions.

\subsection{Spatio-Temporal and Real-time Methods}

For real-time recognition of human actions, spatio-temporal methods can provide accurate performance.  Such methods sacrifice fine details of the movement in order to provide a more computationally efficient solution.  There are many spatio-temporal methods, and the term is used as an umbrella phrase for a wide class of implementations.  Nonetheless, such methods share a common theme - they capture global spatio-temporal characteristics from \textit{optical flow}. Several spatio-temporal approaches have been explored, specific to the human motion recognition problem. Some authors compared surfaces traced out in time \citep{Blank2005}, while others seek representations based upon moments,  \citep{Achard2008} or \citep{Bobick01}.   

The basic idea is that different actions can be distinguished by their unique spatio-temporal flow fields.  By capturing the information from these flow fields,  highly discriminatory feature vectors for time in the video can be constructed.  Comparing different actions is tantamount to comparing these feature vectors.  Because such comparisons are efficient, these methods are attractive for multimedia annotation and querying \citep{Ren2009}.   The  feature vectors can be inserted directly as metadata at each point within a video shot file, and/or as information in the database, to be used in future queries. It is in this way that the video search is performed by \textit{content} and not only with semantic keywords.  Therefore, a video shot query consists of comparing its set of spatio-temporal vectors - each representing segments of the video along the timeline - with the corresponding feature vector sets, stored as metadata within all videos of the database. 

An elegant way to capture the spatio-temporal characteristics of some motion in a video is to transform the original image sequence into a simplified set of images,  called \emph{motion templates}.  These templates provide a quantized representation of the original image frames determined from a background/foreground segmentation technique, such as frame differencing. For example, given a video shot of some human motion, one set of motion templates may be the binarized (i.e., only black/white) human silhouette obtained from pairwise frame differencing during the action. Classic work using spatio-temporal templates for video shots was first described in  \citep{Bobick96}.  In that work, the authors developed motion templates based upon frame difference information. In particular, they introduced the concept of the MHI (motion history instance) and the MEI (motion energy instance) and demonstrated their ability to distinguish people from their gait. Later, \citep{VenkateshBabu2004} used similar motion templates to distinguish  different types of  human actions in video sequences.

For motion templates based upon frame differencing, robust algorithms that eliminate background noise are critical. Several frame differencing algorithms have been proposed that perform interpolation and smoothing by using strong features, such as SIFT, and reducing uncorrelated differences between images.  The resulting difference vectors, when represented on a grid are referred to as \emph{dense optical flow}. One implementation, available in the popular library OpenCV, uses a polynomial technique \citep{Farn03} to optimize the input parameters for obtaining the best foreground optical flow for situations with complex scenes and potentially consisting of moving backgrounds.  Another useful implementation in OpenCV is the  multi-scale pyramid Lucas-Kanade algorithm \citep{Lucas1981} for selecting the scale of the optical flow. 

We recently described a new template, the Motion Vector Flow Instance (MVFI) \citep{Olivieri2012, Pino2014} that utilizes a  dense optical flow algorithm for encoding both the magnitude and direction of the foreground motion.  This encoding scheme improves the discrimination results of human movements from previously employed motion templates, because it contains both first and second derivatives of the velocity field.    As described, a dimensionality reduction transformation is applied to the image sequence after subtracting the mean motion.  The insight of this step can also be found in face recognition, with the concept of eigenfaces \citep{Etemad1997}, where better discrimination and separability of different face classes are achieved by projecting along principal components derived from differences of all images from the mean, called the covariance matrix.   Thus, in the same way that the essential features of a face are the same, but what is important in distinguishing two faces are the slight difference of facial features or the covariance;  the same is true in human motion.  

The covariance eigenspace transformation for human movement was first described in \citep{Huang99B}  to  distinguish the way people walk (their gait).   By using supervised learning with PCA and Fisher Linear Discriminant Analysis  (LDA), they classified gait of different people by pre-assigning the projections of images of a video shot into the training eigenspace.  The Fisher LDA simultaneously minimizes the in-class variance (same actions are closer) while maximizes the out of class variance, thereby separating different classes in the space.  Other studies, such as \citep{Lam2007},  applied this technique in order to identify general  human actions in different  environments.   More recently,  \citep{Cho09} used this PCA+LDA technique to analyze the gait from a set of subjects in order to establish a quantitative grading that could be useful for diagnosing the level of Parkinson disease. 

The PCA method  finds the linear eigenspace transformation for a given dataset that has the maximum projection of the data along the new basis vectors. However, the PCA is a linear transformation, meaning that the orthogonal space is obtained through a combination of translations and rotations of the original space.   The KPCA extends this idea to include nonlinear transformations with the use of the \emph{kernel trick}  \citep{Bishop2006, Mohri2012}. The choice of the kernel function provides the ability to fine tune the solution space for a given input.  As should be expected, the linear PCA solution can be recovered from the KPCA method by choosing a constant kernel function. Recently, \citep{Ekinci2007} used the KPCA approach for gait recognition, and others have applied this technique for improving face recognition \citep{Luh2011,Xie2006}.

\section{Spatio-Temporal Trajectories}

In this section, we provide the technical details behind our spatio-temporal classification method summarized in Figure \ref{fig:workflow}.  We implemented our system and algorithms in Python and make use of several libraries including Scipy/Numpy and OpenCV (ver2.4),  an well-known open source library for computer vision. We also developed a graphical interface in PyQT (QT4 library extensions for python) and produce real-time 3-dimensional plots of the spatio-temporal with MayaVi. 

\subsection{The MVFI spatio-temporal template}

\begin{figure}[ht]
\begin{center}
\epsfig{file=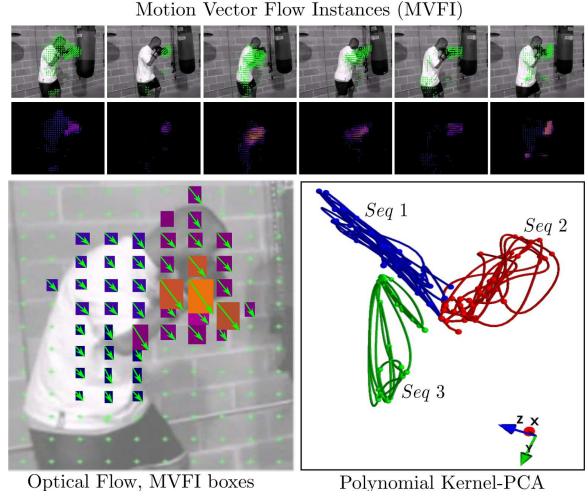, scale=0.58, angle=0}
\caption{(a) The steps for constructing a MVFI template from dense optical flow field. (b) An example frame showing the optical flow field (b-1), the corresponding MVFI encoding (b-2), and the final grayscale MVFI image frame (b-3) }
\label{fig:mvfi}
\end{center}
\end{figure}

In \citep{Olivieri2012}, we described the MVFI (Motion Vector Flow Instance) spatio-temporal template that encodes the velocity field of different human movements. These templates are formed by obtaining a representation of the optical flow field,  $\mathbf{f}$,  of the foreground motion on an evenly spaced grid that are mapped on each image frame at $(x_n,y_m)$. From this flow field, boxes sizes encode the direction while the pixel color encodes the velocity magnitude.   For an input video consisting of $N_f$ frames, this procedure will produce a corresponding video sequence of template frames having the $N_{f -1}$ frames.  A summary of the steps are illustrated in Figure \ref{fig:mvfi}.

Figure \ref{fig:mvfi}(b-1) illustrates the basic idea of how the MVFI is constructed with a boxing video shot.  Using a particular frame in the video sequence, the optical flow vectors are superimposed image.  The algorithm uses this information to create a template, consisting of boxes whose size and shape represents the direction of the vector, and the pixel intensity, an indication of the relative strength of the vector.  The construction proceeds as follows:  an empty storage list $l$, used as a temporary container for manipulating vectors $\mathbf{f}$ at time $t$. For each optical flow grid point $(x_i,y_j)$, information about the vector is used to form boxes that are pushed onto the list $l$.  Next, this list $l$ is sorted by box size so that the largest box is on top.   To construct the final image templates at each time, $t$, the boxes pushed off the sorted list $l$ and drawn within an empty image frame.  In this way, the template accentuates the largest velocity components placing these vectors on top, which will be visible in the template sequence.  This same procedure is repeated for all subsequent image frames in the video shot. 

We showed in \citep{Olivieri2012} that velocity information improves the recognition performance of human actions over previous methods, since these templates capture an instantaneous snapshot of the entire velocity field.  Rapid velocity changes, relative to the mean velocity, will have corresponding trajectories that are very far from the origin of the canonical KPCA space.  In this way, such trajectories are easily distinguished from human movement with small velocity components. Because most human actions are well differentiated by the velocity of body parts and full body movement, these templates are particularly effective for discriminating different types of such actions. 

\subsection{Mathematics of the PCA and KPCA space}

We refer to the spatio-temporal template sequence of human actions as $s$, where there is one template image for each frame in the original video shot.   A particular template image in the sequence is given by $\mathbf{x}_i$, and there are $N_s$ such image templates in the sequence $s$.     

For the purpose of supervised learning, there will be $N_c$ video shots for a particular human action class, $c$.  For training, we combine all image templates $\mathbf{x}_i$  from all the video shots into one column vector, $\mathbf{X} = \mathbf{x}_1 \cdots \mathbf{x}_{s_1}, \cdots \mathbf{x}_{s_c}$.   Thus, $\mathbf{x}_{i,j}$, an element of $\mathbf{X}$,  is an  image template pertaining to the $i$th class, and having the $j$th frame within the sequence $s$.  The total number of images in $\mathbf{X}$ is $N_T$, which is given by the sum $N_T = \sum_i^c N_{c_i}$. The training set, is given by the vector $\mathbf{X}=[\mathbf{x}_{1,1}\cdots \mathbf{x}_{1,N_1}, \mathbf{x}_{2,1} \cdots \mathbf{x}_{2,N_2},\cdots,\mathbf{x}_{c,1}\cdots \mathbf{x}_{c,N_c} ]$,  where each $\mathbf{x}_{i,j}$ is a matrix of the pixels in the image frame  $(i,j)$.  The training vector $\mathbf{X}$ is a column vector consisting of all the pixels from the image sequence. 

\paragraph{Linear PCA}
 
This space is constructed from the orthogonal vectors that possess the most variance between all the images in $\mathbf{X}$.  A reduced dimensional PCA space is found by first obtaining the mean of the vector $\mathbf{X}$, given by $\mathbf{m_x}$, and then obtaining the covariance,  $\mathbf{C_x}$, representing pixels that deviate from the mean: 
\begin{gather*}
 \mathbf{m_x} = \frac{1}{N_T} \sum_{i=1}^c\sum_{j=1}^{N_{c_i}} \mathbf{x}_{i,j} \\
 \mathbf{C_x} = \frac{1}{N_T} \sum_{i=1}^c \sum_{j=1}^{N_{c_i}}(x_{ij}-m_x)(x_{ij} - m_x )^T 
\end{gather*}
The matrix $\mathbf{C_x}$ is found by calculating the contribution from all pixels relative to this mean,  $\bar{\mathbf{X}} = (\mathbf{X}  - \mathbf{m_x})$, so that, 
\[  \mathbf{C_x} = \frac{1}{N} \bar{\mathbf{X}}\bar{\mathbf{X}}^T \]
The orthogonal directions with the most variance are found from the eigenvectors $\mathbf{u}_i$ and eigenvalues $\lambda_i$ of $\mathbf{C_x}$: 
\[ \mathbf{C_x}\mathbf{u}_i = \lambda_i \mathbf{u}_i \]
assuming that $\mathbf{C_x}$ can be diagonalized.  However, $\bar{\mathbf{X}}\bar{\mathbf{X}}^T$ is a very large $n\times n$ 
matrix ($n$ is the total number of pixels of $\bar{\mathbf{X}}$).  

In practice, this excessively large matrix above is simplified \citep{Fukunaga1990} with the relation $\tilde{\mathbf{C}}_x = \bar{\mathbf{X}}^T\bar{\mathbf{X}}$, which is a smaller matrix (only $N_T \times N_T$) amenable to diagonalization.  From this modified eigenvalue equation, the set of eigenvectors and eigenvalues ($\tilde{\mathbf{u}}_i, \tilde{\lambda}_i$) that span the space of $\tilde{\mathbf{C}}_x$ are approximately equivalent to those of the original matrix,  $\mathbf{C_x}$, thereby justifying the truncation of the matrix.

A further approximation is made to reduce the solution spectrum to a small number of eigenvectors. Such an approximation is justified since the values of the eigenvalues $\lambda_j$ decrease monotonically fast for modest eigenvectors indices, $j$, so that  $\lambda_j \approx 0$ for $j>k$. Thus, the $K-$dimensional eigenspace is truncated so that only the $k\leq K$ largest eigenvalues $|\lambda_1| \geq |\lambda_2| \geq \cdots |\lambda_k|$ are kept.  In practice, we truncate the basis at $K=10$.  The partial set of eigenvectors span a space $\mathbf{y} = [ \mathbf{y}_{1,1}\cdots \mathbf{y}_{c,N_c}]$, and represent projections of the original images: 
\[ \mathbf{y}_{i,j} = [ \mathbf{u}_1 \cdots 
             \mathbf{u}_k]^T \mathbf{x}_{i,j}  = \mathbf{E} \mathbf{x}_{i,j}\]   

The above mathematical procedure describes the precise manner by which the image sequence is converted to a spatio-temporal trajectory; each point representing  one template in this reduced dimensional eigenspace.

\paragraph{The KPCA}

The PCA is a linear rotation of the original $d$-dimensional bases into one having maximum variance for the given data set.  Intuitively, if the data were a general ellipsoid having some angle with respect to the original axes, the PCA transformation would discover the rotation coincident with the principal axes of the ellipsoid. Such linear transformations may not be optimal and a more general nonlinear transformation could provide a better solution.  The KPCA method \citep{Scholkopf1999}, retains the concept of PCA, but can be nonlinear.   The method uses the \textit{kernel trick} - that states that only the form of the inner product needs to be specified, not the bases functions, making it a practical method implement. In practice, an appropriate kernel $K(\mathbf{x},\mathbf{x^\prime})$ is chosen with model parameters adjusted that maximize the \textit{out-of-class} separation while minimize the \textit{in-class} separation. 

The detailed mathematics for constructing the kernel-PCA method can be found elsewhere \citep{Bishop2006}, however we describe briefly its use for obtaining spatio-temporal trajectories.  As before, we construct column vector with all  the template images from the video shots in the training set:   $\mathbf{\bar{X}}$ (with $N_T$ elements).   Also, as before, we subtract the mean movement from $\mathbf{X}$.  A nonlinear transformation that will reduce the space to an $M$-dimensional space is found by postulating bases vectors $\phi_i(\mathbf{x})$, so that each point is projected onto these directions $\phi_i(\mathbf{x})$, where $\sum_{n}\phi(\mathbf{\bar{x}}_n)=0$ (with $n=1,\cdots,N_T$)

The $M \times M$ covariance matrix is given by:  
\[ \mathbf{C_{\bar{X}}}=\frac{1}{N_T}\sum_{n=1}^{N_T}\phi(\mathbf{\bar{x}}_n)\phi(\mathbf{\bar{x}}_n)^T \]

An appropriate solution  eigenvalue problem: 
\[ \mathbf{C_{\bar{X}}}\mathbf{u^\prime}_i=\lambda_i^\prime\mathbf{u^\prime}_i \]
is found by diagonalizing $\mathbf{C}_{\bar{X}}$

After algebraic manipulations, the \textit{kernel trick} consists in finding a transformation where only form of the inner product is needed to project the original vector into this newly postulated space having basis vectors $\phi(\mathbf{x})$.   In this way,  the form of the eigenvectors $\phi$ do not need to be calculated in order to find projections.    Instead we write the transformation in terms of the inner product, here called the kernel function, given by $k(\mathbf{x},\mathbf{x_n})$.

A projection of the original point into this space along the $i$th component is written as: 
\[
\mathbf{y}_i(\mathbf{x})=\phi(\mathbf{x})^T\mathbf{u^\prime}_i=\sum_{n=1}^{N_T}a_{in}k(\mathbf{x},\mathbf{x}_n)
\]
where $a_{in}$ are the coefficients for each eigenvector that are obtained based on the normalization condition. 
\[ 1=\lambda^\prime_i N_T \mathbf{a}_i^T \mathbf{a}_i \]
We use a polynomial kernel with an optimized order of the 
polynomial $d$ for the analyzed data:
\[ k(\mathbf{x},\mathbf{y}) = (\mathbf{x}^T\mathbf{y} + c)^d\]

\subsection{Recognition of new actions from a KNN distance of trajectories}

With the  kernel-PCA transformation from a  training set, we can classify a query video by projecting it into this newly formed space and comparing it to the trajectories corresponding to the training set. The exact procedure is as follows: a query video containing a human action is processed with low level image processing algorithms to create the set of  MVFI templates.  These templates are then projected into the newly formed space through a KPCA transformation.   A distance metric, such as KNN, could be used  to calculate the proximity of constituent points along the  trajectory into each of the defined classes.  Depending upon a pre-established threshold,  the query video shot is classified depending upon the percentage of points pertaining to each class. 

We used the public KTH database \citep{Schuldt2004} for performing training and validation of our algorithm.  In particular, we performed tests with the following six actions: walking, jogging, running, boxing, clapping and waving.   From our own human action database, we also studied four  actions: jogging, boxing, playing tennis and greeting. 

\begin{figure}[ht]
\begin{center}
\epsfig{file=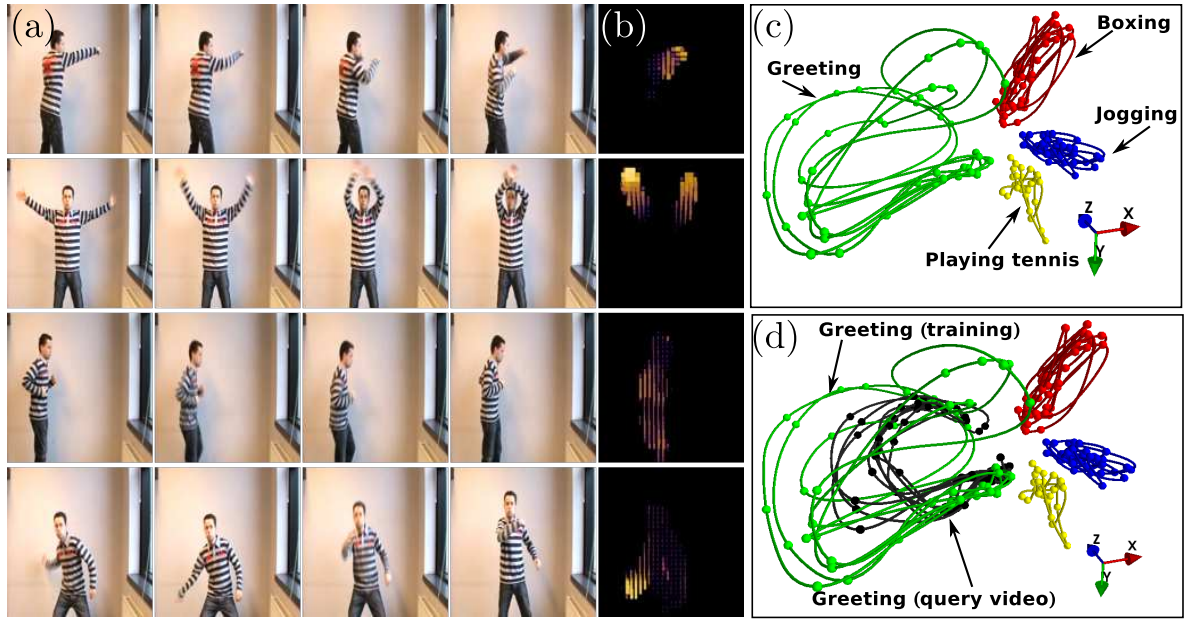, scale=0.5, angle=0}
\caption{(a) The polynomial KPCA space from training with a set of  video shots consisting of the actions:  boxing, greeting, jogging and playing tennis.  (b) the projection of a query video sequence (greeting) into the space trained by a video shots of (a).}
\label{fig:projection}
\end{center}
\end{figure}

Figure \ref{fig:projection}(a) shows the spatio-temporal trajectories, or projections,  into the polynomial KPCA eigenspace constructed from four different human actions.  The set of kernel parameters were selected that provide maximum separation of the four classes.  Query video shots containing one of the trained actions were transformed and projected into the space for classification, as shown in Figure \ref{fig:projection}b.  As can be seen,  the trajectory is closest to those trajectories corresponding to the same action.  By calculating the KNN distance between the query shot and the trajectories stored for each video in the database along its timeline, similarity scores were obtained.    

%---------------------------------
\section{Using local differential curvature for distinguishing action classes}
\label{sec:sec3}

One of the problems with the traditional KNN distance metric, as described in the previous section,  for distinguishing different action classes from the spatio-temporal trajectories is the ambiguities for points along the curve that cross into different class boundaries, especially near the origin.   We will recall points near the origin in the covariance space represent parts of the motion having small velocity components. Many actions can have at least some parts during their motion with small velocity, so the overlap with another action class in the space is common.  Thus, a distance metric solely based on the Euclidean separation between points or groups, loses information about the  \emph{connectedness} and \emph{spatial orientation} of the full trajectory curve.

Instead, we define a new concept of points along the trajectory, the \textit{trajectory point cloud} that allows us to define a new distance metric based upon the local differential geometry of the curve. This new method uses different scales of the human action spatio-temporal trajectories.   Viewed from far away, the spatio-temporal curves lie within unique \emph{mean (osculating)  hyperplanes}.   By determining the hyperplane of different trajectories, we can distinguish the different corresponding actions.  On a finer scale, each point has local geometric characteristics, such as the curvature and torsion, providing information about how it is connected in time.   We can use this local information to provide better KNN class discrimination at  a finer scale.   Thus, we shall define a distance metric that combines the knowledge from different scales to classify trajectories.   We call this classification, the \emph{trajectory point cloud classifier}. 

We can find the mean  hyperplane from local properties of the curve.  A qualitative description of our method is as follows.  The spatio-temporal trajectory is parameterized by a constant speed arc length, simplifying the differential geometry.  We divide this trajectory into sequential segments, $\gamma_i$ (where $i=1, \cdots t_s$), that overlap in a way similar to a moving window. We use these segments to determine the local properties of the curve:  the curvature, torsion and the co-moving orthogonal basis along the arc length, from the generalized $n$-dimensional Frenet-Serret (FS) equations.  For each segment $\gamma_i$, we obtain its so-called binormal vector $\mathbf{b}_i$, which defines the \emph{osculating} plane traced out by this curve.  By summing the weighted contribution of all such binormal vectors the $\mathbf{b}_i$, we obtain the  \emph{mean osculating hyperplane}  for the entire trajectory.   Each binormal vector is weighted by a term proportional to the radius of curvature. Recall that the curvature is a measure of how much the curve deviates from a straight line, while  the torsion is a measure how much the curve moves out of the plane.   Thus,  those segments with large radii of curvature contribute the most in defining the mean hyperplane, while those that tightly closed, having a high curvature, contribute less.  The unique hyperplane can be used in a distance metric to distinguish different trajectories based upon the angles between the trajectory planes.

The \emph{trajectory point cloud} $\mathcal{C}$ is a way of describing the different scales associated with the trajectory.  Locally, each trajectory point contains not only its spatial position, but how it is connected to other points.   At a larger scale, the entire trajectory can be treated as a cloud of points, having a centroid $\mathcal{C}_\text{cm}$ and mean radius.   Therefore, this multi-scale information is used to distinguish trajectories in three situations related to the separation between cloud centroids, namely when it is (a) approximately zero (overlapping clouds), (b) approximately the radius of a cloud, or c) larger than several cloud radii.  The first and last (a and  c) are classified well with clustering methods, such as the easily implemented KNN.  For the case when trajectories overlap, however, we can use additional information of the local geometric properties to distinguish points.  With our new geometric formalism of trajectories, we treat this in two ways:  with mean osculating hyperplane orientations and to distinguish finer details, with a fuzzy-KNN like method. 

\subsection{Definitions of the trajectory point cloud}

\begin{figure*}[htb]
\begin{center}  
\epsfig{file=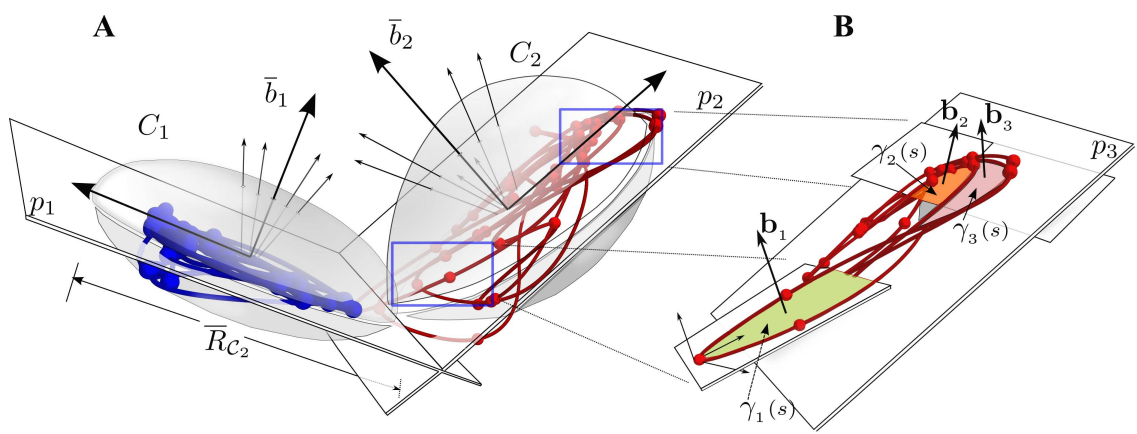, width=0.9\textwidth, angle=0}
\caption{A. Illustration of different hyperplanes formed from the average spatio-temporal \emph{trajectory point cloud}.  The spatio-temporal trajectories trace out curves where average osculating plane can define different classes. B. Detailed view of three sequential overlapping curve segments,  $\gamma_i$, showing the binormal vectors to their osculating planes.}
\label{fig:TrajectoryCloud}   
\end{center}
\end{figure*}

To aid in the definitions and concepts, Figure \ref{fig:TrajectoryCloud}A  shows the trajectories from two different human actions and the associated \emph{trajectory point clouds}, $\mathcal{C}_1$, and $\mathcal{C}_2$.  The points along the trajectories represent the MVFI image templates transformed into the KPCA space.  Two characteristics are evident upon visual inspection:  the curves appear to lie in separate planes, and they are partially overlapping.  The figure shows the cloud surface;  the mean cloud radius $\overline{R}_{\mathcal{C}}$, which is used for the distance metric.  The vectors $\overline{\mathbf{b}_i}$ are the resultant weighted normal vectors to the \emph{time averaged hyperplane}.

Figure \ref{fig:TrajectoryCloud}B  shows two isolated regions along the trajectory, while all other details and parts of the trajectory have been removed for visual clarity.  In these isolated regions, particular discrete curve segments, $\gamma_i$ have been selected out for illustration.  In the algorithm, these sequential overlapping curve segments form a set  $\{\gamma_i \}$, as described above.  The figure shows how each curve segment, $\gamma_i$, can be used to calculate the FS local frame, the curvature and torsion.  While each segment define slightly different planes and have different curvature, the aggregate will define an average plane for the entire trajectory.   In Figure \ref{fig:TrajectoryCloud},  $\gamma_1(s)$ is the segment in green, with the binormal vector $\mathbf{b}_1$, that is slightly out of the plane defined by $\gamma_3(s)$. The segment defined by $\gamma_2$ is also out of plane and has binormal vector $\mathbf{b}_2$.   Since the curvature of $\gamma_2$ will be higher than $\gamma_1$ or $\gamma_3$, it will contribute less to the resultant vector, since we calculate this resulting vector weighted by the radius of curvature.

These concepts are illustrated further in Figure \ref{fig:weightedSum}A, which also serves to define the variables involved. Two segments, $\gamma_1$ and $\gamma_2$,  of a single trajectory are represented.  Segment $\gamma_1$ lies within the plane $p_1$, while $\gamma_2$ lies within plane $p_2$.  For each, we can make the following definitions.

\subsection{Local Differential properties}

\paragraph{The Trajectory}

We now formalize the ideas described previously.  A trajectory curve, $\mathbf{r}$, is parameterized by the arc length $s$ through the mapping $s \mapsto \mathbf{r}(s)$.   In practice, represent the spatio-temporal trajectory as in terms of a $\mathcal{B}$-splines.  $\mathcal{B}$-splines are smooth functions and parameterized in terms of the arc-length.  In this way, they can be used to calculate local differential properties of the trajectories in a practical and numerically efficient way.   

Formally, we can write the $k$th degree $\mathcal{B}$-spline, $S_{k,t}(x)$ and its first derivative as: 
\begin{align}
 S_{k,t}(x)       &= \sum_i \alpha_i \mathcal{B}_{i,k,t}(x) \\
 S^\prime_{k,t}    &=  \frac{d}{dx} \sum_i \alpha_i \mathcal{B}_{i,k} = \sum_i (k-1) \frac{\alpha_i - \alpha_{i-1}}{t_{i+k-1} - t_i} \mathcal{B}_{i,k-1}  
\end{align}
where $\mathcal{B}_{i,k,t}(s)$  are  piecewise polynomial basis functions that are functions of the arc length $s$.   The $k$ points are called knots and are the control points along the arc length of the curve.  Higher order derivatives can be obtained in a similar way.  These equations are used to obtain polynomial expressions for the curvature $\kappa(s)$, the torsion $\tau(s)$, and the Frenet-Serret basis vectors.  

\paragraph{Arc segment and local frame}

We define a discrete segments of arc along the $k-th$ trajectory $\mathbf{r}_k(s)$ as $\gamma_i^k$ with length $\sigma_i = \Delta s$. Thus, the trajectory consists of a collection of such arc segments: $\mathbf{r}_k(s) =  \{ \gamma_i(\sigma_i) \}$,  for $i=1, 2, \cdots, n$. In practice we take the arc lengths to be equal so that $\sigma_i = \sigma$ for all $i$. 

For each segment $\gamma_i(\sigma)$, we can calculate the average curvature $\kappa_i(\sigma)$ and torsion $\tau_i(\sigma)$ centered within the interval $\sigma$ at $s_i$, by integrating over the arc segment.   The equations of the curvature $\kappa(s)$ and torsion $\tau(s)$ in terms of the trajectory $\mathbf{r}(s)$ along the entire curve, and the mean values for each segment $\gamma_i$ are given by: 

\begin{align*}
\kappa(s)  & = \frac{ | \mathbf{r}^\prime(s) \times   \mathbf{r}^{\prime\prime}(s)|}{ | \mathbf{r}^\prime(s)|^3 }, \quad
& \langle \kappa(s_i) \rangle  =  \frac{1}{\sigma} \int_{\gamma_i}  \kappa(s^\prime) ds^\prime \\
\tau(s)    & = \frac{ ( \mathbf{r}^\prime(s) \times   \mathbf{r}^{\prime\prime}(s)) \cdot  \mathbf{r}^{\prime\prime\prime}(s) }{   | \mathbf{r}^\prime(s) \times   \mathbf{r}^{\prime\prime}(s)|^2    }, \quad 
& \langle \tau(s_i) \rangle  =  \frac{1}{\sigma} \int_{\gamma_i}  \tau(s^\prime) ds^\prime 
\end{align*}

\begin{figure*}[htb]
\begin{center}  
\epsfig{file=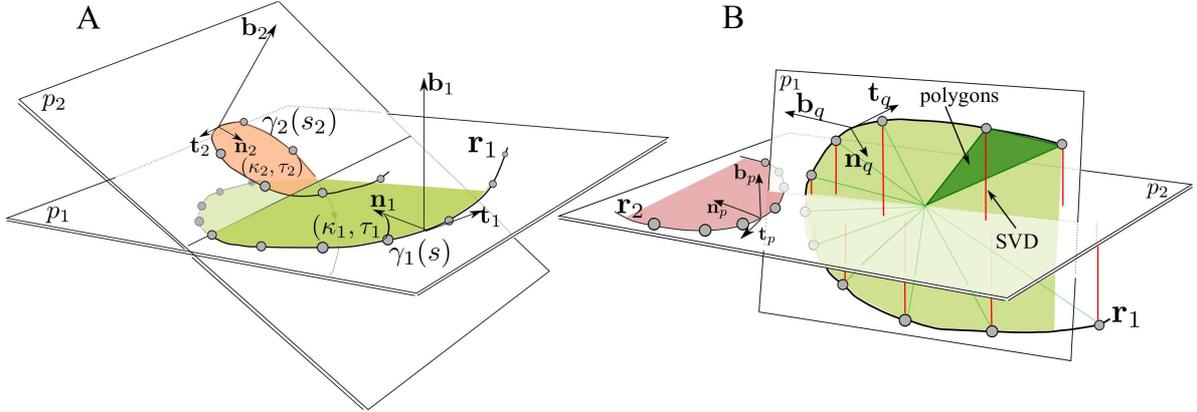, width=0.95\textwidth, angle=0}
\caption{A.  Weighted sum of Binormal vectors.  The FS frame for each curve provides a weighted direction based upon the mean radius of curvature and the torsion. B.  Comparison showing how the FS binormal vector uses the trajectory to be able to determine the correct plane as opposed to a SVD method (red lines) or purely geometric triangulation (green lines).}
\label{fig:weightedSum}   
\end{center}
\end{figure*}

With these quantities, we can obtain the local basis frame from the general $n$-dimensional Frenet-Serret (FS) equations, given in terms of the vectors $(\mathbf{t}, \mathbf{n}, \mathbf{b})$, well-known from the theory of curves.   The \emph{tangential vector} is the derivative of the trajectory with respect to the arc length $s$,  $\mathbf{t} = d\mathbf{r}/ds = \mathbf{r}^\prime$ (Figure \ref{fig:weightedSum}A, $\mathbf{t}_1$ is tangent to the curve $\gamma_1$ ). The \emph{normal vector} is found by taking the derivative with respect to the $\mathbf{t}$ and inversely proportional to the curvature.  Thus, $\mathbf{n} =  \mathbf{t}^\prime/|| \mathbf{t}^\prime || = \mathbf{t}^\prime/\kappa(s)$.   The \emph{binormal vector}  is found by taking the cross product between the normal and tangential vector and also related to  the torsion:  $\mathbf{b}  =  \mathbf{t} \times \mathbf{n}$.   Now we have the exact equation of the binormal vector that is used to define the plane of the curve, or the \emph{osculating plane}.

Given an entire trajectory $\mathbf{r}(s) = \{ \gamma_i \}$, we can find the FS frame, mean curvature, and mean torsion for each segment $\gamma_{i,k}$, so that $\xi_i=(\mathbf{t}_i, \mathbf{n}_i, \mathbf{b}_i, \langle\kappa_i\rangle, \langle\tau_i\rangle )_k$.   The mean osculating plane can be found by summing the \emph{weighted} contributions $\mathbf{b}_i$  from all arc segments, and the resulting vector $\langle b_k \rangle$ defines the plane for the $k-$th trajectory.  We can see in Figure \ref{fig:weightedSum}A  how the weighted contribution of each $\mathbf{b}_i$  depends upon the curvature.  In particular,  small tightly curved loops (large $\kappa$), indicated by $\gamma_2$, should contribute less in defining the mean plane than large radius segments (shown as $\gamma_1$ in the figure).  Making connection to the temporal dependence of the trajectory as points in a video sequence, the resultant binormal vector  is really a \emph{time averaged osculating plane}.  The equation is given by: 
\[   \langle\mathbf{b}\rangle  = \beta \sum_i \biggl ( \frac{\beta}{\kappa^2_i}  \biggr )  \mathbf{b}_i \]  
where $\beta$ is a normalization constant.

\paragraph{Alternatives descriptions of planes}

Many alternative techniques exists for obtaining the mean hyperplane that cuts through a set of points, that need not rely upon the differential properties of curves.  Nonetheless, the method we developed has the advantage of providing local geometric information that can be used on several scales.  Figure \ref{fig:weightedSum}B illustrates two alternative methods for obtaining a mean plane through a set of points in 3-dimensions.  If no knowledge is available for how points are connected, Singular Value Decomposition (SVD) provides a simple projection procedure for finding the best fit plane through points in a least square sense.  This method will often fail to coincide with the plane defined by connected points, as shown in Figure Figure \ref{fig:weightedSum}B (indicated by plane $p_2$).    A method for obtaining a mean plane from connected points is to construct successive polygon segments, also shown in Figure \ref{fig:weightedSum}B  (and later in Figure \ref{fig:PlanesCompare}).   This method yields the same plane as that defined by the binormal vector. In this method,  however, all other quantities must still be calculated for other steps in our the classification algorithm.  

\subsection{Steps in the trajectory point cloud algorithm}

\begin{figure*}[htb]
\begin{center}  
\epsfig{file=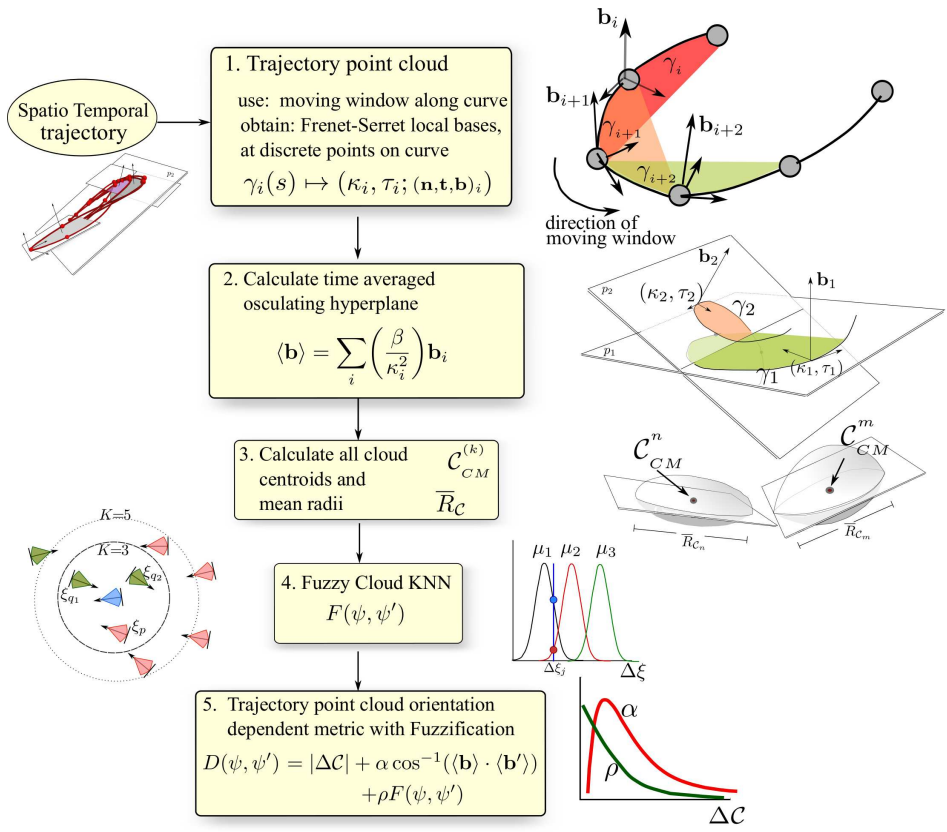, width=0.75\textwidth, angle=0}
\caption{Steps in the algorithm for inferring the action classes based upon the concept of trajectory point cloud. }
\label{fig:distAlgorithm}   
\end{center}
\end{figure*}

The steps of the the \emph{trajectory point cloud classifier} algorithm are shown in Figure \ref{fig:distAlgorithm}.   In the previous section,  we described steps 1 and 2, where we defined the concept of the trajectory point cloud with the collection of segments, $\gamma_i$, and the time averaged osculating plane from the resulting binormal vector $\langle\mathbf{b}\rangle$.   We now use this information to develop a distance metric that classify an unknown video into a one of a set of trained classes. 

In step 3 of \ref{fig:distAlgorithm}), we use each trajectory $\mathbf{r}_k$ to calculate macroscopic quantities: the cloud centroid from the trajectory $\mathbf{r}_{\text{CM}}^{(k)}$, and the as well as the average \textit{cloud radius} $\overline{R}_{\mathcal{C}_k}$. From these definitions, we can express each trajectory cloud $\mathcal{C}^k$  as the tuple:   
\[  \psi_k \equiv \psi(\mathcal{C}^k) = (\{\xi_i\}^k , \langle \mathbf{b}_k\rangle, \mathbf{r}_{\text{CM}}^{(k)},  \overline{R}_{\mathcal{C}_k}) \] 
where the set $\{\xi_i\} =(\mathbf{t}_i, \mathbf{n}_i, \mathbf{b}_i, \langle\kappa_i\rangle, \langle\tau_i\rangle )_k$,  are the local properties of each trajectory point in the cloud.

\begin{figure*}[htb]
\begin{center}  
\epsfig{file=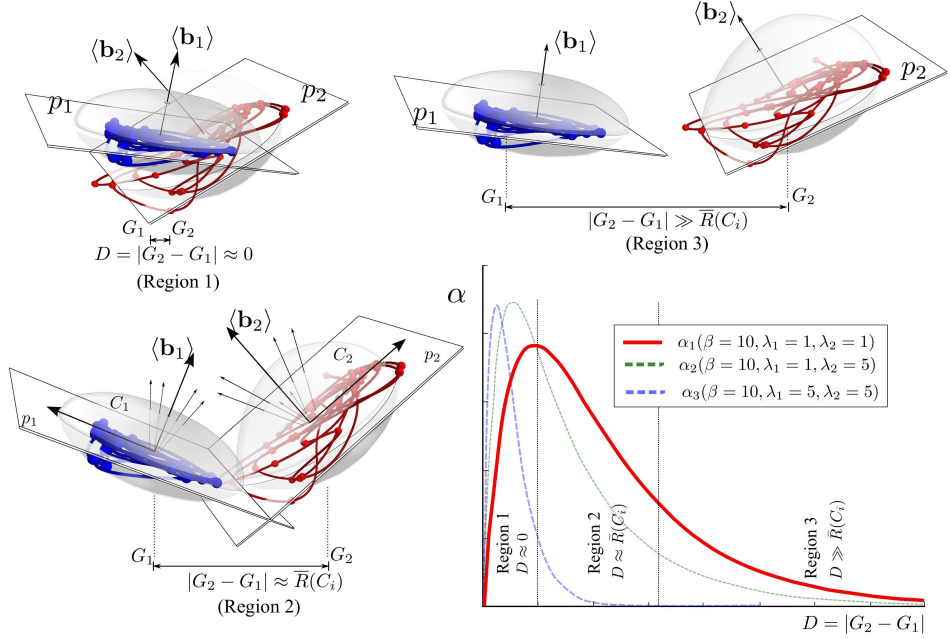, width=0.75\textwidth, angle=0}
\caption{(Bottom right) The modulation function, $\alpha$, for the \emph{hyperplane orientation} term in the distance metric between trajectory point clouds.   (Top left, right; bottom left)  The three different scenarios which define that regions of the function $\alpha$.}
\label{fig:distanceMetric}   
\end{center}
\end{figure*}

How does this trajectory information help to distinguish between different  action classes?  Figure \ref{fig:distanceMetric} illustrates different configuration scenarios that can occur with respect to the trajectory point clouds. The configurations define three separate regions that our distance metric will be selectively sensitive:  
\begin{itemize}
\item \emph{Region 1} (top left):  when the trajectories overlap. This is the case where the trajectories correspond to the same action. For this situation,  we want the distance to only depend upon the centroid ($\mathbf{r}_{\text{CM}}^{(k)}$), which is close to zero. Thus, we want to eliminate contributions of the distance metric that correspond to the orientation of the mean hyperplanes.   If we wish to distinguish fine details between actions of the same class, we will use a specialized KNN, we call the fuzzy cloud KNN, briefly described below.
\item \emph{Region 2} (bottom left): This is when the trajectories are separated by at least a mean cloud radius, $\overline{R}_\mathcal{C}$.  In this case, the trajectories can be partially overlapping.   This is precisely the region where ambiguities can arise in other metrics.  Here we see the power of the hyperplane method.  In this case, we want the contribution to the distance metric from the  hyperplane normal vectors $\langle\mathbf{b}_k\rangle$  to be maximum.  
\item \emph{Region 3} (top right): This is the case when the trajectories are separated larger than a few cloud radii.  In this case, the cloud centroid is sufficient to resolve different classes.  Thus, here the contribution from the hyperplane orientation should also be decreasingly small as the separation distance between the cloud centroid is increased.
\end{itemize}

These ideas are captured in the function $\alpha$ (bottom right) as a function of the trajectory point cloud separation  $\Delta C_{\text{cm}}(\psi, \psi^\prime)$.  The function can treat the three regions above in a different manner: \textit{(a)}  it is zero  when the separation is approximately zero $\alpha(\Delta C_{\text{cm}}\approx 0) \rightarrow 0$,  \textit{(b)}  it is maximum when the separation is a mean radius, $\alpha (\Delta C_{\text{cm}}(\psi, \psi^\prime) \approx \overline{R}) =  \alpha_max$ , and \textit{(c)}  it decreases exponentially for separations greater than a mean radius,   $\alpha (\Delta C_{\text{cm}}(\psi, \psi^\prime) \gg \overline{R}) \sim \exp ( - \lambda \Delta C_{\text{cm}}(\psi, \psi^\prime)) $

The function  $\alpha$  that will modulate the \emph{hyperplane orientation} in the distance metric between trajectory point clouds is shown in Figure \ref{fig:distanceMetric}(bottom right) and is given by:  
\begin{equation}
\alpha= \beta \exp[ - \lambda_1 \Delta  C_{\text{cm}}(\psi, \psi^\prime) ]  \biggl ( \frac{1-\exp ( - \lambda_2 \Delta  C_{\text{cm}}(\psi, \psi^\prime) )  }{\Delta  C_{\text{cm}}(\psi, \psi^\prime)}  \biggr )
\label{eq:alpha}
\end{equation}
where $\psi$ and $\psi^\prime$ are two trajectory cloud tuples defined previously, the free parameters $\beta$, $\lambda_1$, and $\lambda_2$ are chosen as a function of the cloud radius;  $\beta$ is a scaling constant,  $\lambda_1$ controls how steep the function is close to the origin, that is how quickly the function cuts off, while $\lambda_2$ controls the long exponential tail, so that larger values will go to zero faster.   Different values of these parameters are shown in Figure \ref{fig:distanceMetric} in order to illustrate the effect of each of the free parameters.  Values of these parameters for real trajectories of our study are given in below in the experimental results section.

\subsection{The Fuzzy Cloud KNN}

Our trajectory cloud classifier was designed so that when trajectories overlap, the hyperplane orientation can be used to distinguish different actions. However, in some situation, two different actions could have similar hyperplanes. Also, in another situation, we may wish to distinguish the difference between two executions of the same action, as in our recent work that studies the quality of Olympic gymnastics movements \citep{Pino2014}.  For these situations, we can use the set of local trajectory segments, $\xi_i$ to obtain a distance measure.  We developed a specialized KNN algorithm to classify a query trajectory into a set of classes, called the \emph{fuzzy cloud KNN}, that uses the local information of the trajectory.

\begin{figure*}[htb]
\begin{center}
\epsfig{file=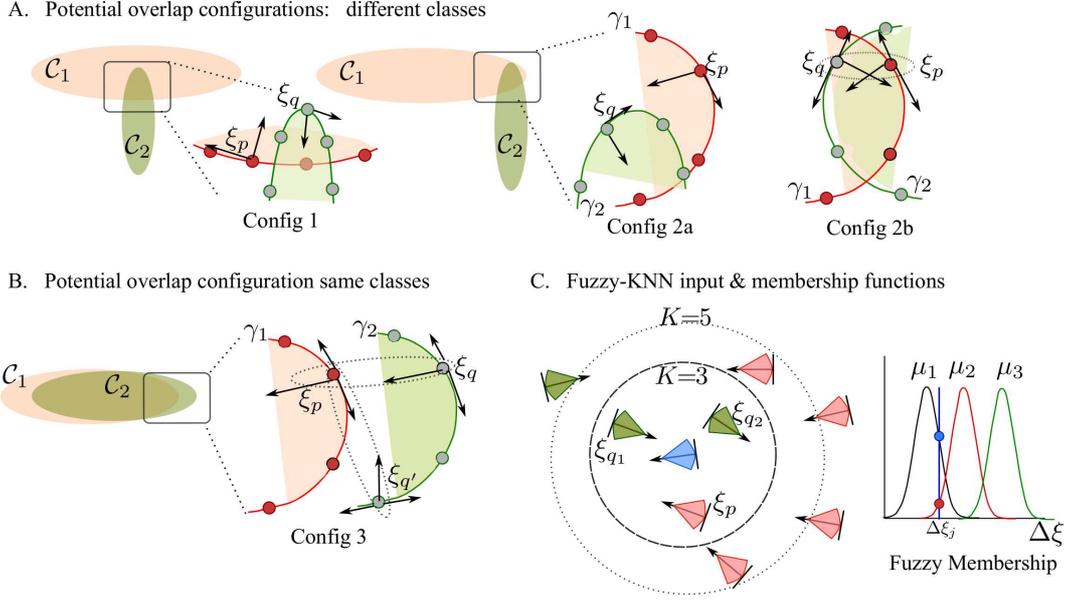, width=0.85\textwidth, angle=0}
\caption{A. The different possible configurations. B: Membership rules based upon the different angles that the FS frame can take on as well as local curvature. C. The membership functions}
\label{fig:FuzzyCloud} 
\end{center}
\end{figure*}

Although the details are beyond the scope of this paper and shall be described elsewhere, Figure \ref{fig:FuzzyCloud} illustrates the general idea of the algorithm.  Different possible overlap configurations are shown in Figure \ref{fig:FuzzyCloud}A and B. The points pertaining to different trajectories are given in different colors and labeled with their trajectory tuple, $\xi_p$ and $\xi_q$, respectively. The situations illustrated in the figure provide the logic for assigning membership rules. In the configurations of type A when clouds overlap at some angle, the normal vector orientations are opposite and the curvatures are large and small, respectively. In configuration B when clouds are nearly coincident, the local trajectory points will have vectors and curvatures that will coincide on average.

Figure \ref{fig:FuzzyCloud}C illustrates the idea of the \emph{fuzzy cloud KNN} using trajectory points, represented as \emph{wedges} to accentuate the orientation $\xi_i$. In the example, a test wedge (shown at the center in blue) is to be classified into either one of two groups (indicated by red and green). Analogous to the classic KNN algorithm, a $K$ value is chosen that determines the maximum nearest neighbors to be considered for the classification of the test point.  As in the original fuzzy-KNN algorithm described by \citet{Keller1985}, these neighboring points are weighted by a set of fuzzy membership functions that are inversely proportional to the separation. In our algorithm, such functions are parameterized by the relative difference, $\Delta \xi_{p,q}=\xi_p-\xi_q$, between the test point and a neighboring point (pertaining to one of the classes), and written as $\mu_j(\Delta\xi)$, with the quantization $j=1,\dots m$.  Rather than assigning crisp class membership for the test point, this procedure produces a set of vectors whose components are the values of $\mu_j(p,q)$, between $\xi_p$ and $\xi_q$.  These vectors are used in an aggregate function $F(\xi, \xi^\prime)$, which defines a set of rules for class inference.

\subsection{The Distance Metric}

Given the above definitions, we can now define the full distance metric between trajectory point clouds, which consist of three terms: one that depends on the centroid distance, another that depends upon the orientation of the hyperplanes, and another that can provide fine structure details from a fuzzy KNN like inference:  
\begin{equation}
 D( \psi, \psi^\prime )  = \overbrace{ |\Delta C_{\text{cm}}(\psi, \psi^\prime)| }^\text{CM distance} +  \underbrace{  \alpha  \cos^{-1}\biggl (   \langle\mathbf{b}_k\rangle  \cdot  \langle\mathbf{b}_{k^\prime}\rangle  \biggr  )}_\text{plane orientation dependent} 
+ \overbrace{\rho F(\psi, \psi^\prime) }^\text{fuzzyKNN}  
\label{eq:distanceMetric}
\end{equation}
where $\Delta\mathcal{C}_\text{CM} = \mathcal{C}_{\text{CM}}^{(k)} - \mathcal{C}_{\text{CM}}^{(k^\prime)}$,  $\alpha$  modulates the strength of the hyperplane orientation (as shown in Figure \ref{fig:distanceMetric}), and $\rho$ modulates the strength of the fuzzy cloud KNN penalty function $F(\psi, \psi^\prime)$, so that it contributes when the  trajectory clouds partially or fully overlap.  
Since the function $F(\psi, \psi^\prime)$ produces solutions that depend on the class type, this function contributes differently to $D( \psi, \psi^\prime )$ for each class.

\subsection{Implementation and Results of the TPC Classifier}

We implemented the formalism for the \emph{trajectory point cloud} (TPC) in a set of Python classes that depend only upon Numpy/Scipy/Matplotlib libraries for numerical operations and plotting.  The B-spline routine from \verb|scipy.interpolate| was used to represent the curves and higher order derivatives.  All other functions were implemented given the descriptions provided in previous sections.  

\begin{figure*}[htb]
\begin{center}  
\epsfig{file=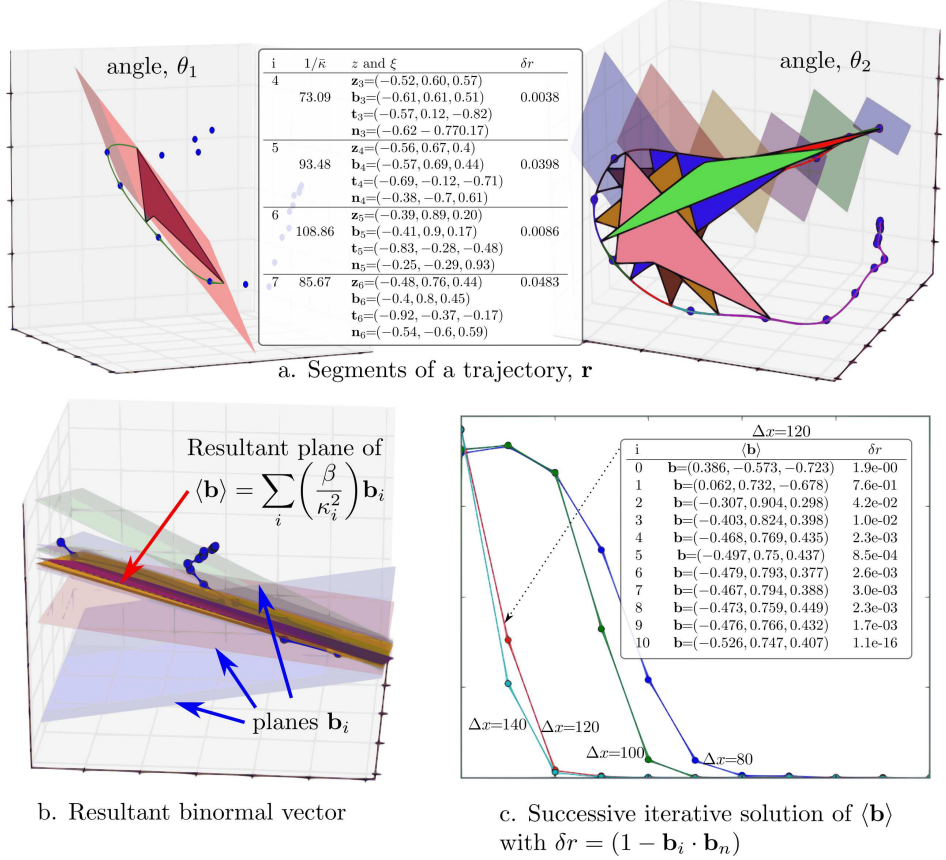, width=0.75\textwidth, angle=0}
\caption{The results of obtaining the binormal vectors $\mathbf{b}_i$ along the trajectory and the resultant binormal vector $\langle\mathbf{b}\rangle$.}
\label{fig:BinormalVec}   
\end{center}
\end{figure*}

Figure \ref{fig:BinormalVec} shows the results of calculating the binormal vector for a particular spatio-temporal trajectory.   In particular, Figure \ref{fig:BinormalVec}a (left) shows a plane obtained with the binormal vectors $\mathbf{b}_i$ for each individual segments $\gamma_i$ and the correspondence with the polygon plane for the same segment.  Figure \ref{fig:BinormalVec}a (right) shows the same trajectory with many other $\gamma_i$ segments and corresponding planes defined by $\mathbf{b}_i$.  The values of the arc-length averaged radii of curvature, normal vector to polygon (given by $\mathbf{z}$) and FS vectors ($\mathbf{b}, \mathbf{n}, \mathbf{t}$) are given in the table inset.   As seen, the binormal vectors are coincident with the polygon normal vectors. Figure \ref{fig:BinormalVec}b, shows successive solution by summing each $\mathbf{b}_i$ along the trajectory.  The  are drawn resultant binormal vector $\langle\mathbf{b}\rangle$ is indicated in the figure by the darkest plane and indicated by the arrow.   Figure \ref{fig:BinormalVec}c shows the convergence of $\langle\mathbf{b}\rangle$ with successive addition of each segments $\gamma_i$ for different values of the segment length $\Delta s$.

\begin{figure*}[htb]
\begin{center}  
\epsfig{file=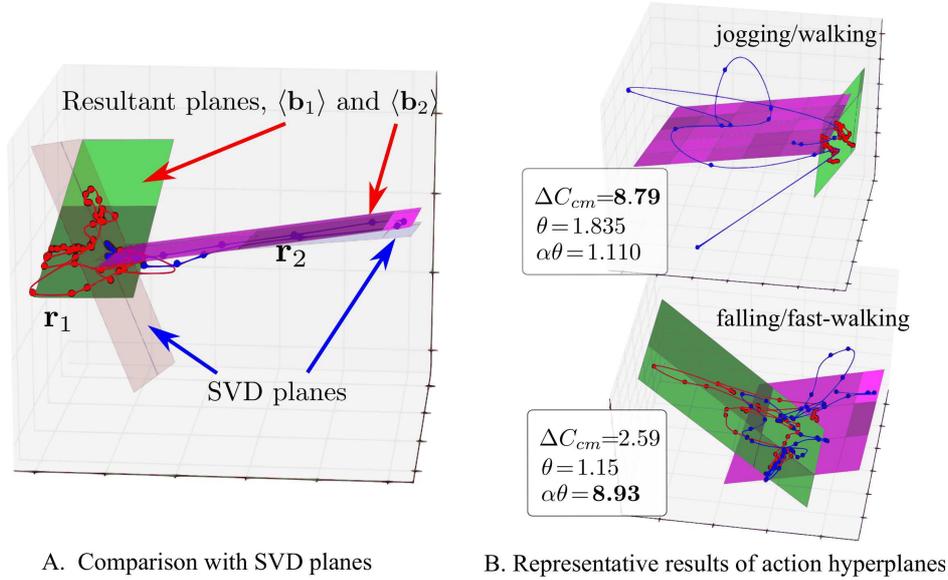, width=0.75\textwidth, angle=0}
\caption{A. Comparison of mean planes of two trajectories calculated with two different methods, SVD and resultant binormal vector.  B. Representative results showing the two terms involved in the distance metric calculation.  The angle $\theta=cos^{-1}(\mathbf{b}\cdot\mathbf{b}^\prime)$ is the angle  between planes. }
\label{fig:PlanesCompare}   
\end{center}
\end{figure*}

Figure \ref{fig:PlanesCompare}a, shows planes for the trajectories of two actions.  For comparison, planes were calculated from the SVD method and the resultant binormal vector method described above.  For the case of trajectory $\mathbf{r}_1$ (right), both methods are  similar.  However, in the case of the trajectory $\mathbf{r}_2$, the SVD fails to properly calculate the plane for the closed connected curve, while the mean binormal plane is correct.  

Figure \ref{fig:PlanesCompare}b shows two separate action comparisons that can suffer from ambiguities with the classical KNN:  (top) jogging/walking, and (bottom) falling/fast-walking.   Since the covariance space is different depending upon the actions trained,  we normalized all quantities with respect to the separation maximum extent of the two clouds $\langle R_1 + R_2\rangle$.    In the modulation function $\alpha$ (Equation \ref{eq:alpha}), we set empirically  $\lambda_1=2.5$ and $\lambda_2=25.0$, in order to peak close to $\Delta C_{\text{cm}}\approx 0.2$ and have a long tail, guaranteeing a contribution from the hyperplane orientation term for trajectories that are relatively close, while moderate for those further away.    As can be seen from the values,  the distance metric for the jogging/walking case (top) is dominated by the first term of Eq. \ref{eq:distanceMetric} (having a value of $\Delta C=8.79$, while the second term is $\alpha\theta=1.1$),   while the falling/fast-walking  case (bottom) is dominated by the second term of Eq. \ref{eq:distanceMetric} ($\Delta C$ is less than $\alpha\theta$)  which depends on the angle between hyperplanes.

\section{Experimental Results of CBVR}

From the spatio-temporal analysis with a KPCA and our new \textit{trajectory point cloud} classifier described in the previous sections, we validated the recognition performance of our CBVR system using two public video datasets (MILE database \citep{Olivieri2012}) and (KTH database \citep{Schuldt2004}).  Another objective of these tests was to show that well chosen parameters in a KPCA can outperform the recognition rates of a linear-PCA, while still retaining computational performance.  For this, we fine-tuned the polynomial kernel function of the KPCA in order to  maximize the class separation of human activities in the study.  

\subsection{Experiments in MILE video database}

The specifics of our database \citep{Olivieri2012} are as follows.   It consists of  240 video shot sequences representing 4 human actions (boxing, greeting, playing tennis and jogging) recorded with 12 different people. The video shots were obtained under normal lighting conditions using a commodity Sony (DCR-HC15) MiniDV, with a sampling rate of 25 frames/s. All actions were recorded using the same focal distance and no special backlighting preparations were implemented. The videos were saved in AVI MPEG encoding format. Together with the raw footage, we processed each video shot with an adaptive resizing algorithm to create image sequences of $256 \times 256$, for later use in our CBVR system. Figure \ref{fig:projection} shows a sampling of different MVFI templates (b) in BGR color space that result from the different human actions (a).  Finally, the frames are converted to grayscale for vector quantization of the spatio-temporal templates.

We carried out experimental tests with a training set consisting of 64 video shots (8 people, 4 human actions, and 2 video shots for each person): boxing ($c_1$), greeting ($c_2$), jogging ($c_3$) and playing tennis ($c_4$).   For controls in our analysis, we also considered two cases: (1) a \emph{null} action, defined as a scene without a human action, and (2) a \emph{non-defined} action, which are other actions not considered in the training set.  In the case of a \emph{null} action, the resulting  trajectories in the PCA eigenspace are concentrated close to the origin.

\begin{figure}[ht]
\begin{center}
\epsfig{file=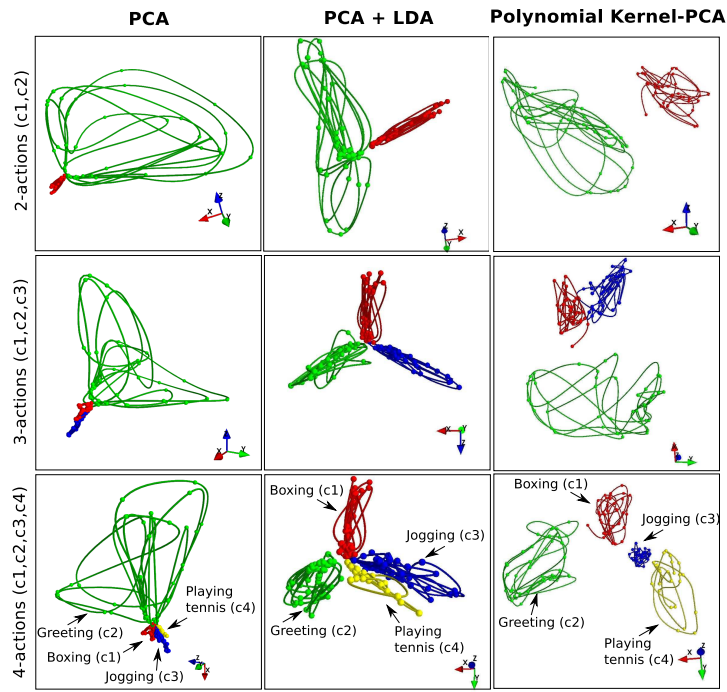, scale=0.61, angle=0}
\caption{Comparison of the PCA, PCA+LDA and polynomial KPCA spaces with 2-class, 3-class and 4-class training cases.}
\label{fig:comparison}
\end{center}
\end{figure}

Figure \ref{fig:comparison} shows a comparison of both the linear and polynomial kernel PCA applied to one of the four classes training discussed above.  The example shows the spaces formed with  two, three and four separate human action classes,  each represented by a single video shot and a single person.  The results demonstrate that we can achieve a  better separation between the different classes from the KPCA, than can be obtained from the linear-PCA.   Indeed, by fine tuning the kernel function parameters, we can control the class separation, which ultimately can lead to improved classification performance of the algorithm.

\begin{figure}[htb]
\begin{center}
\epsfig{file=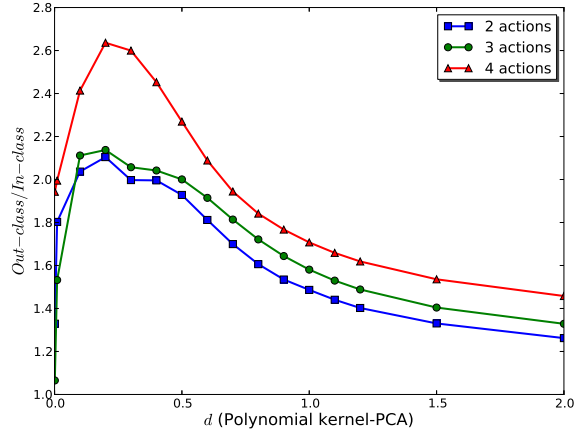, scale=0.38, angle=0}
\caption{Selection of the polynomial kernel parameter $d$ for maximum class separation. The difference  between \emph{out-of-class} and \emph{in-class} is highlighted.}
\label{fig:polynomial}
\end{center}
\end{figure}

The polynomial kernel takes the form: 
\[ k(x,y) = (x^Ty)^d\] 
where the value of  $d$ is selected to maximize the class separation.  The dependence this parameter on the class separation is shown in Figure \ref{fig:polynomial} that shows action classification results for different values of $d$.   Just as in the Fisher criteria, the objective function seeks a constrained maximization solution:  maximizing the average distance between points of the eigenspace trajectory belonging different classes (\textit{out-class}) while minimizing the average distances among points belonging to the same class (\textit{in-class}). These relations are shown in Figure \ref{fig:polynomial} for plots of the ratio of \textit{out-class} and  \textit{in-class}, corresponding to training data previously shown in Figure \ref{fig:comparison}. These studies indicate that the optimal value of the tuning parameter, $d$, is independent of  the human action type as well as total number of classes in the training set. 

\subsection{Results of trajectory point cloud classifier}

\begin{figure}[ht]
\begin{center}
\epsfig{file=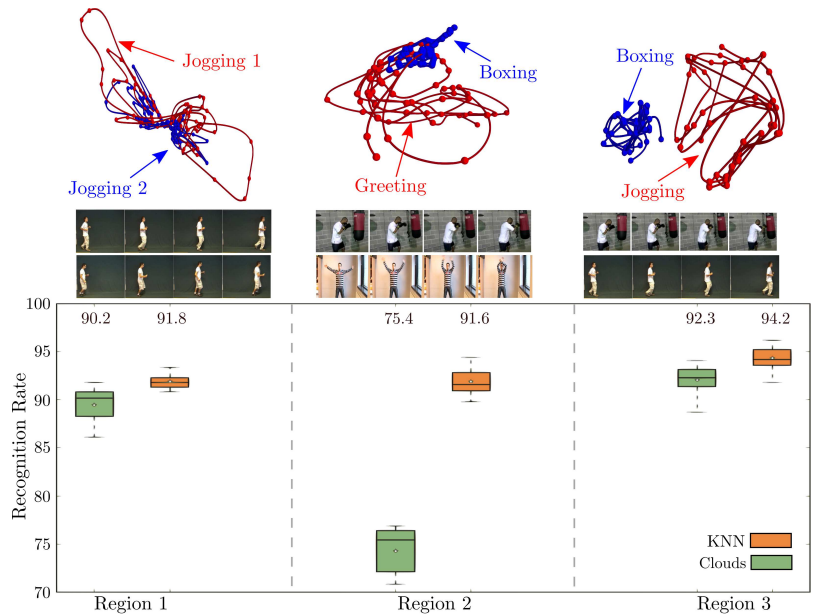, scale=0.38, angle=0}
\caption{Results of Cloud classifier compared to KNN.}
\label{fig:cloudClassify}
\end{center}
\end{figure}

Figure \ref{fig:cloudClassify} shows the results of the trajectory point cloud classifier. 

In order to quantify the distance between classes, we used a simple Euclidean metric. Thus, given two points $\mathbf{p}_i$ and $\mathbf{q}_j$, in separate classes, $i$ and $j$ respectively, within the $n$-dimensional space, the distance $d_{i,j} = (\mathbf{p}_i, \mathbf{q}_j)= |\mathbf{p}_i - \mathbf{q}_j| $. A metric for the total distance between classes $i$ and $j$ is to sum all pairwise distances $D=\sum_{i\ne j} d_{i,j}$. 

We compared the PCA and kernel-PCA methods by normalizing the distance vectors obtained in the respective spaces, dividing by the largest distances: $\max\{D_{Kernel}^i \}$ or $\max\{D_{PCA}^i\}$, along the principal axes, $\mathbf{e}_j$. From these maximum values, we defined the ratio $r = \max\{D_{Kernel}^i / \max\{D_{PCA}^i\}$ and the normalization factor $F$, such that: 
\[F = 
   \begin{cases}
        r    & \max\{D_{Kernel}^i \} > \max\{D_{PCA}^i\}\\ 
        1/r  & \max\{D_{Kernel}^i \} < \max\{D_{PCA}^i\},
        \end{cases}.
\]

The result of this normalization procedure is shown in  Figure \ref{fig:cloudClassify}, consisting of the results obtained from the distances between the classes indicated in Figure \ref{fig:comparison}. In all cases,  the polynomial kernel-PCA provided superior class separation and recognition results,  even when the linear PCA is combined with linear discriminant analysis (LDA).

\subsection{Experiments in KTH database}

\begin{figure}[htb]
\begin{center}
\epsfig{file=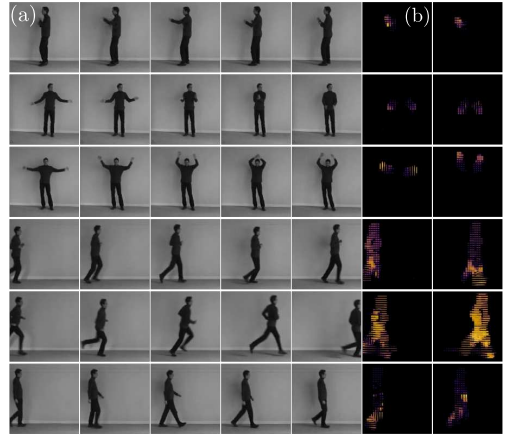, scale=.34, angle=0}
\caption{Human actions used from KTH dataset.}
\label{fig:kth}
\end{center}
\end{figure}

The KTH video database \citep{Schuldt2004} is a widely used public databases for testing and comparing human motion recognition algorithms. This database contains six action classes  (boxing, hand clapping, hand waving, jogging, running and walking).  These actions were recorded with 25 people in four different scenarios (figure \ref{fig:kth}a): $s_1$ \emph{outdoors}, the camera is parallel to the object moving trajectories, $s_2$ \emph{outdoors}, there is an angle between the camera and the object moving trajectories, or there are scale changes, $s_3$ \emph{outdoors}, there are different clothes or  pack on the back, and  $s_4$ indoors,  there are various degrees of shadows. Figure \ref{fig:kth} (b) illustrates an example of MVFI templates a sampling of video frames from this database. 

From the KTH database, the \emph{training set} we selected consists of  six human actions performed by eight different people. All the other videos in the database were used as the test set.  The results of the recognition performance is given in the confusion matrix of Table \ref{tab:confusionMatrix}. The confusion matrix provides a comparison between the results obtained with the linear PCA and KPCA. The  lowest recognition rate corresponds to the running actions, given the similarity with jogging in the database.  As in previous comparisons,  the polynomial KPCA provided better discrimination amongst the different actions when  compared with the linear PCA.

\begin{table}[htb]
\caption{The confusion matrix from results using the KTH database. The recognition rates were obtained with the \textit{trajectory point cloud} hyperplanes classifier (without brackets) and with KNN classifier (in brackets). }
\centering
\footnotesize
\label{tab:confusionMatrix}
  \begin{tabular}{| c | c  c  c  c  c  c  c |}
  \hline
  \multirow{7}{2mm}{\begin{sideways}{\textbf{PCA}}\end{sideways}}&
   & \textbf{Box} & \textbf{Clap} & \textbf{Wave} & \textbf{Jog} 
	 & \textbf{Run} & \textbf{Walk} \\ \cline{2-8}
   & \multirow{1}{*}{\textbf{Box}} & \textbf{91.2 (89.5)}   & 6.9 (7.6) & 1.9 (2.9) & 0 & 0 & 0 \\
   & \multirow{1}{*}{\textbf{Clap}} & 9.6 (12.3)& \textbf{84.3 (79.8)} & 6.1 (7.9)& 0 & 0 & 0 \\
   & \multirow{1}{*}{\textbf{Wave}} & 4.3 (5.2) & 10.1 (11.3) & \textbf{85.6 (83.5)} & 0 & 0 & 0 \\
   & \multirow{1}{*}{\textbf{Jog}} & 0 & 0 & 0 & \textbf{91.8 (89.6)} & 6.7 (8.8)& 1.5 (1.6)\\
   & \multirow{1}{*}{\textbf{Run}} & 0 & 0 & 0 & 14.1 (17.9) & \textbf{83.8 (77.3)} & 2.1 (4.8)\\
   & \multirow{1}{*}{\textbf{Walk}} & 0 & 0 & 0 & 5.2 (4.7) & 1.2 (2.1) & \textbf{93.6 (93.2)} \\ \cline{1-8}
           
     \multirow{7}{2mm}{\begin{sideways}{\textbf{PCA+LDA}}\end{sideways}}&
   & \textbf{Box} & \textbf{Clap} & \textbf{Wave} & \textbf{Jog} & \textbf{Run} 
   & \textbf{Walk} \\ \cline{2-8}
   & \multirow{1}{*}{\textbf{Box}} & \textbf{93.7 (90.2)}  & 5.4 (8.3) & 0.9 (1.5) & 0 & 0 & 0 \\
   & \multirow{1}{*}{\textbf{Clap}} & 6.4 (9.6) & \textbf{91.1 (87.3)} & 2.5 (3.1)& 0 & 0 & 0 \\
   & \multirow{1}{*}{\textbf{Wave}} & 3.9 (3.6)  & 5.7 (5.8)& \textbf{90.4 (90.6)} & 0 & 0 & 0 \\
   & \multirow{1}{*}{\textbf{Jog}} & 0 & 0 & 0 & \textbf{93.6 (91.4)} & 5.0 (6.5)& 1.4 (2.1)\\
   & \multirow{1}{*}{\textbf{Run}} & 0 & 0 & 0 & 8.4 (9.5)& \textbf{89.7 (85.2)} & 1.9 (2.1)\\
   & \multirow{1}{*}{\textbf{Walk}} & 0 & 0 & 0 & 7.8 (7.6)& 0.2 (0.3) & \textbf{92 (92.1)} \\ \cline{1-8}
  
    \multirow{7}{2mm}{\begin{sideways}{\textbf{Pol. kernel-PCA}}\end{sideways}}&
   & \textbf{Box} & \textbf{Clap} & \textbf{Wave} & \textbf{Jog} 
	 & \textbf{Run} & \textbf{Walk} \\ \cline{2-8}
   & \multirow{1}{*}{\textbf{Box}} & \textbf{94.6 (92.4)}  & 3.8 (5.1)& 1.6 (2.5)& 0 & 0 & 0 \\
   & \multirow{1}{*}{\textbf{Clap}} & 5.7 (7.8) & \textbf{93.1 (89.2)} & 1.2 (3.0)& 0 & 0 & 0 \\
   & \multirow{1}{*}{\textbf{Wave}} & 1.1 (2.4) & 5.2 (6.8) & \textbf{93.7 (90.8)} & 0 & 0 & 0 \\
   & \multirow{1}{*}{\textbf{Jog}} & 0 & 0 & 0 & \textbf{94.8 (92.6)} & 3.7 (5.1) & 1.5 (2.3)\\
   & \multirow{1}{*}{\textbf{Run}} & 0 & 0 & 0 & 6.4 (8.7)& \textbf{92.3 (88.8)} & 1.3 (2.5)\\
   & \multirow{1}{*}{\textbf{Walk}} & 0 & 0 & 0 & 3.9 (5.1)& 1.0 (0.8)& \textbf{95.1 (94.1)} \\ \cline{1-8}
   
  \end{tabular}
\end{table}

The average recognition rate is a useful metric for comparing the performance of different classifiers for human actions.  Table \ref{tab:comparison} shows the average recognition rate from our results and compared with results published previously by other researchers. Our
results,  using the MVFI templates with either the PCA and KPCA,  outperformed other recognition techniques.  Our system achieves real-time recognition with an accuracy greater than 93\%.   From the details provided from other published results, we could not determine if the techniques function in real-time or not. 

\newcommand{\minitab}[4][l]{\begin{tabular}{#1}#2\end{tabular}}
\begin{table}[htb]
\caption{Comparison of different methods applied to the KTH database. Our methods results are presented with \textit{trajectory point cloud} hyperplanes classifier (without brackets) and with the KNN classifier distances (in brackets). }
\footnotesize
\centering
\label{tab:comparison}
  \begin{tabular}{| l | c |}
  \hline
  \multicolumn{1}{|c|}{Methods} & \multicolumn{1}{c|}{Recognition} \\
	\multicolumn{1}{|c|}{ } & \multicolumn{1}{c|}{accuracy (\%)} \\
  \hline
  \textbf{Pol. kernel-PCA + MVFI (this paper)} & \textbf{93.9 (91.3)} \\
  \textbf{PCA + LDA + MVFI (this paper)} & \textbf{91.8 (89.5)} \\
  \textbf{PCA + MVFI (this paper)} & \textbf{88.4 (85.5)} \\
  Liu and Shah \cite{Liu2008} & 94.2 \\
  Mikolajczyk et al. \cite{Mikolajczyk2008} & 93.2 \\
  Schindler et al. \cite{Schindler2008} & 92.7 \\ 
  Laptev et al. \cite{Laptev2008} & 91.8 \\  
  Jhuang et al. \cite{Jhuang2007} & 91.7 \\
  \hline
  \end{tabular}
\end{table}

\subsection{Experiments as a CBVR: Video indexing and Annotation}

\begin{figure}[ht]
\begin{center}
\epsfig{file=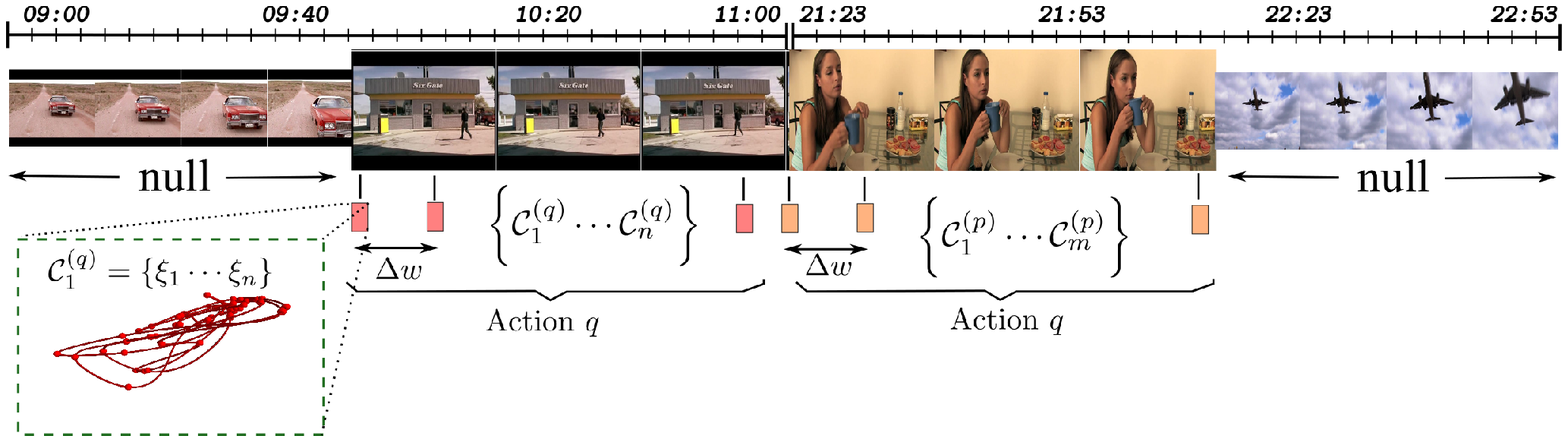, scale=.8, angle=0}
\caption{Bag of Features.  Schematic illustration of how the trajectory point cloud and features are stored along the timeline of the video for indexing and querying. }
\label{fig:exampleAnnotation}
\end{center}

\end{figure}

Once an action is identified, a full video sequence can be annotated, marking those parts of the video containing relevant human actions and possibly storing this information as metadata.   As our algorithm marches through the video, it must decide whether a trained event is present or not. In particular, the routine identifies human action in the training set as well as  \emph{non-actions} or null frames. The algorithm processes $m$ frames of the timeline in a video at a time, performing the KPCA transformation, and calculating the distance metric to trained classes. The algorithm proceeds with an overlapping moving window, of $n$ frames, thereby determining actions for every $n$ frames.  The essential steps in the algorithm are given as follows: 

We used our system to be able to annotate sections of videos and return the time intervals during which large actions take place. We have tested our algorithm with several films to try to identify 5 human actions: picking up the phone, drinking, sitting, walking and running. We contemplate the null case to classify any other action such as  “a car moving” or “a dog playing”.  Figure \ref{fig:exampleAnnotation} illustrates indexing the timeline with trajectory point cloud and associated feature tuple $\xi$.   A query shot will make comparisons to each of these vectors. 

\begin{figure}[ht]
\begin{center}
\epsfig{file=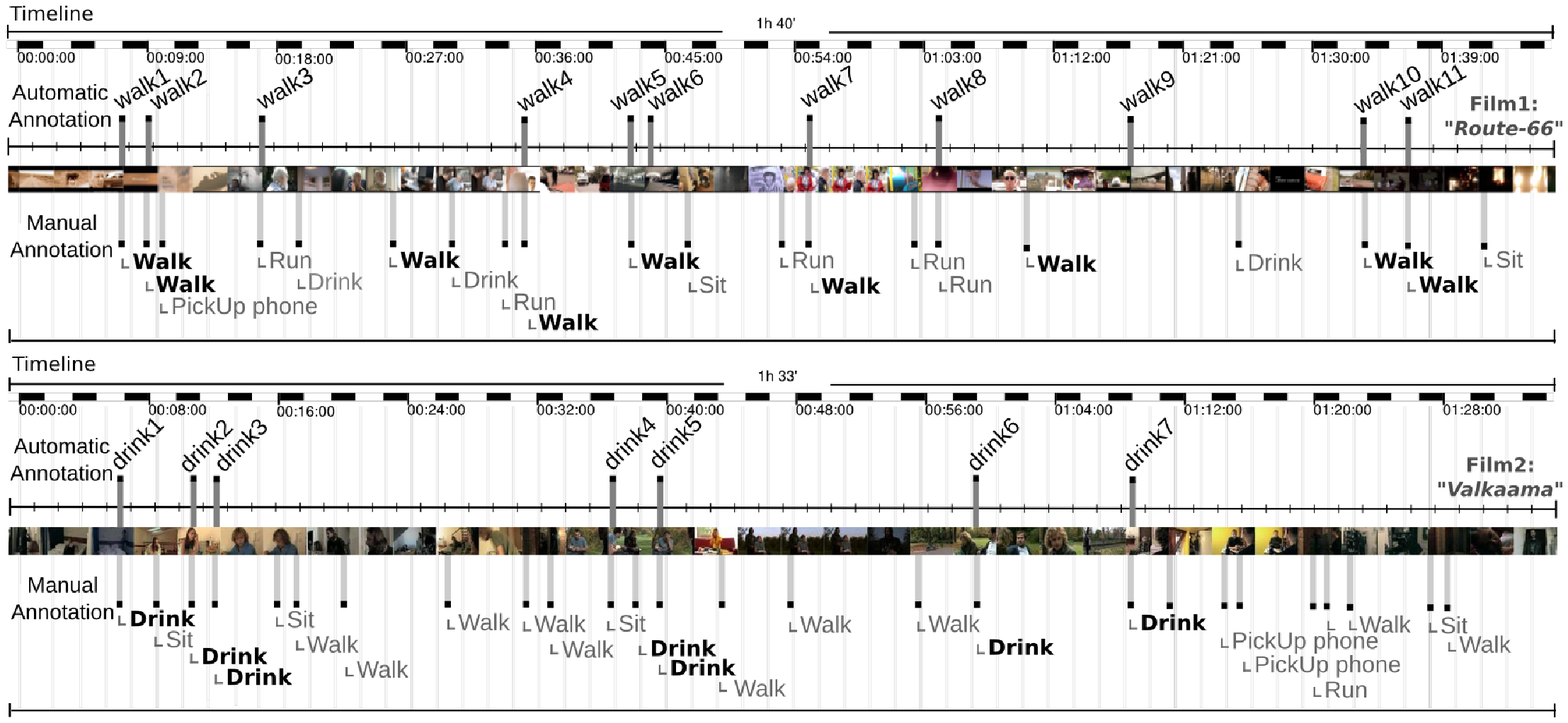, scale=.6, angle=0}
\caption{Results of annotation of six human actions in two films (“Route-66” and “Valkaama”). Shown are the False Negatives (FN) and False Positives (FP) of a binary classification between ‘walk’ and any other action for “Route-66” and between ‘drink’ and any other action for “Valkaama” are shown.}
\label{fig:resultsAnnotation}
 \end{center}
\end{figure}

We performed ground truth validation tests of our algorithm by applying it to two feature length movies that we annotated manually.  From these tests, we determined the recognition rate of  our algorithm for detecting the location of actions similar to those in our training set. Figure \ref{fig:resultsAnnotation} shows the results for detecting  two actions, "walking" and "drinking" in two open source  films ("Route 66 - an american bad dream" and "Valkaama").  To determine the location of these frames we used a marching moving window, with a window size of $m=250$ frames and an overlap of $n=50$ frames.  The training set was taken from the MILE database  by selecting five actions performed by eight separate people. Figure \ref{fig:resultsAnnotation} shows false positives (FP) and false negatives (FN) for classifications (walk/other actions and drink/other actions).  Many FPs were due to different shot angles and body clipping that were not considered in the training set, but produced similar MVFI spatio-temporal trajectories to those of running and walking.   These can be eliminated by more stringent requirements and by increasing the training set to include more shot angles and body clipping scenes similar to those found in the movie.

We compared our results with a ground truth manual annotation of both full featured films in order to obtain quantitative performance information of the our algorithm, such as the sensitivity and specificity. For each of the films in  Figure \ref{fig:exampleAnnotation}, the manual annotation of the 5 types of human actions shown are included in the training and on top the result of our automatic annotation produced by our system. In the case of the first film "Route-66", the figure shows a scene from the film that our system correctly detected correctly a "walking" action shot, while from the movie "Valkaama", we show a particular results of  "drinking" and “picking up” actions. For each of the actions defined in the study, the results of $TPR$ and $TNR$ are shown in the Table \ref{tab:cbvrRes}. The analyses were made by dividing the actions into groups, in the same way as we explained previously for experiments with the MILE/KTH human movement dataset. Each analysis consists of two groups, (1) the action in question (2) any other action not considered in the study.

\newcommand{\ra}[1]{\renewcommand{\arraystretch}{#1}}

\begin{table*}
\centering
\ra{0.7}
\label{tab:cbvrRes}
\begin{tabular}{@{}cccccccccc@{}} 
\toprule
&  \multirow{2}{*}{\textbf{Actions}} & \multirow{2}{*}{\textbf{Real shots}} & 
\multicolumn{6}{c}{\textbf{CBVR results}} & \\
\cmidrule{5-9}
& & &   \textbf{TP} & \textbf{TN} &\textbf{FP} &\textbf{FN} & \textbf{TPR} & \textbf{TNR} &\\  \midrule
& Walk &  42 &  29 & 54 & 13 & 8 & 0.78 & 0.81 \\
& Run &  28 &  21 & 67 & 8 & 7 & 0.75 & 0.89 \\
& PickUp Phone  & 4  & 3 & 90 & 9 & 1 & 0.75 & 0.91 \\
& Drink  & 18 &  13 & 73 & 12 & 5 & 0.72 & 0.86 \\
& Sit  & 11 &  9 & 78 & 14 & 2 & 0.82 & 0.85 \\
\bottomrule
\end{tabular}
\caption{CVBR results from the 5 different actions in 2 movies (``Route-66'' and ``Valkaama'').}
\end{table*}

\subsection{Web application for CBVR with short video shots}

\begin{figure}[ht]
\begin{center}
\epsfig{file=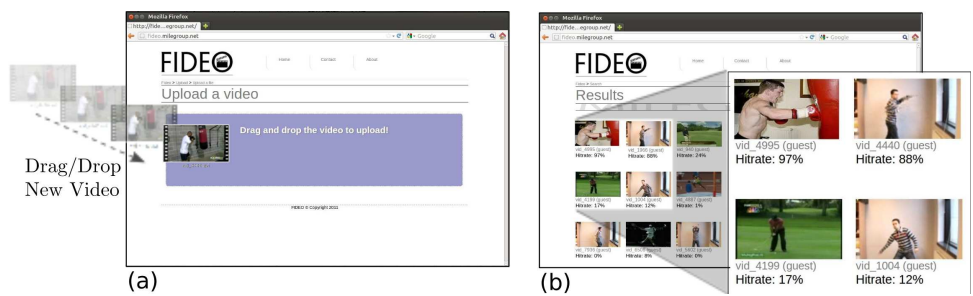, scale=0.65, angle=0}
\caption{(a) Dragging and dropping a video to upload and to analyze.  (b) Result of a query with a video of a person boxing  (application provided at fideo.milegroup.net).}
   \label{fig:query}
 \end{center}
\end{figure}

As an interface for our CBVR algorithms, we developed a lightweight web application (available at \url{http://fideo.milegroup.net}) that can query the database from saved videos using a drag-and-drop search box,  or the query can be made from a live webcam capture of a human action.  An example screenshot of the query-by-video web application is shown in Figure \ref{fig:query}, where results from a query with a boxing action are shown.  In particular, a brief description of the web interface application is described as follows.  For an existing query video,  the shot, is moved into the drag-and-drop search box, uploading the video to the server. For the live stream option, a web application records the video from the webcam and subsequently uploads the result to the server.  Once uploaded, the video shot is processed by a server-side application to produce the corresponding spatio-temporal trajectory that will be used to produce a similarity search against all videos, in the database.   This search is carried out using search windows, so that if a video is longer than the search window, the entire duration of the video is searched to determine the location of the action within the video of the database. 

As described previously, the database server contains the fine-grained spatio-temporal trajectories for all points across their timeline obtained with the KPCA transformation.   For a query shot, similar videos and/or location of video segments in larger videos are found by calculating the pairwise accumulated distance between the targets and the query.  In our example of Figure \ref{fig:query}(b), the recorded video shot correctly produces a higher similarity to all videos with the similar action (boxing), as seen through a higher similarity percentage.   For null actions, or for actions that are not contemplated in the training set, a hit rate should yield negligible hit rate values.

\section{Conclusions}

The spatio-temporal template method allows complex motions to be processed and classified in real-time by using a supervised learning procedure.   We showed that by using a  KPCA transformations,  better out-of-class separation can be obtained by fine tuning the kernel parameter depending upon the nature of the data.  As we postulated, the KPCA provides more flexibility through a nonlinear transformation as compared with the linear-PCA. 

Nonetheless, there is a limit to the extent that different action classes can be separated even with highly tuned kernel engineering of the  KPCA space.  This is especially true as the number of different action classes increases in a multi-class classification analysis.  The scaling to larger classes is accompanied with a commensurate increase in class boundary overlap. As these class boundaries become \emph{softer}, traditional classifiers such as KNN or SVM will be unable to crisply distinguish the membership of certain points in along the trajectory, and therefore, the recognition rate will suffer.  

Thus, the most profound contribution of this paper is a new classifier for spatio-temporal trajectories, that we call the \emph{trajectory point cloud} classifier.   As described, this classifier specifically treats complicated cases but more common case when trajectories partially overlap, namely they are different action classes but there class boundary is not crisp.   Our method considers local differential geometric properties of the trajectories in order to identify the average n-dimensional osculating hyperplane where these trajectories live.   Different actions will lie on hyperplanes that are oriented at different angles and the center of mass of these \emph{trajectory point clouds} will allow us to control the extent to which this orientation is incorporated into the distance calculation between different clouds.   Thus, we say that the distance metric for our classifier is orientation-dependent, and that the direction is determined by the weighted binormal vector to the mean osculating hyperplane obtained by the independent contribution from a collection of sequentially overlapped curve segments along the trajectory. 

Our method resolves the problem of overlapping trajectories as arises more commonly in multi-class analysis.  This is in contrast to traditional methods such as the classical KNN, where the trajectory is  treated as a set of independent points, thereby ignoring essential information about the connectedness of the points.    Thus, we demonstrated that our new \emph{trajectory point cloud} classifier  is superior to the KNN (or other point-centric methods) for detecting human actions with a spatio-temporal methodology.   Nonetheless, even though we described this technique in the context of human motion recognition, the classification technique is general and can be extended to other cases, where the points are correlated, as in this case for time-sequenced video frames.  

Finally, we provided a proof of principle demonstration of how our spatio-temporal MVFI and classification method could be used as a CBVR system to annotate/index and query videos from a multimedia database.   Due to the nearly infinite variety of shot angles and partial body shots, online learning combined with probabilistic inference could cover a wider range of motion variations and contexts.

\footnotesize
\bibliographystyle{elsarticle-harv}
\bibliography{RefKCBVR02}

\begin{thebibliography}{44}
\expandafter\ifx\csname natexlab\endcsname\relax\def\natexlab#1{#1}\fi
\expandafter\ifx\csname url\endcsname\relax
  \def\url#1{\texttt{#1}}\fi
\expandafter\ifx\csname urlprefix\endcsname\relax\def\urlprefix{URL }\fi

\bibitem[{Achard et~al.(2008)Achard, Qu, Mokhber, and Milgram}]{Achard2008}
Achard, C., Qu, X., Mokhber, A., Milgram, M., 2008. A novel approach for
  recognition of human actions with semi-global features. Machine Vision and
  Applications 19~(1), 27--34.

\bibitem[{Beecks et~al.(2010)Beecks, Uysal, and Seidl}]{Beecks2010}
Beecks, C., Uysal, M., Seidl, T., July 2010. A comparative study of similarity
  measures for content-based multimedia retrieval. In: Multimedia and Expo
  (ICME), 2010 IEEE International Conference on. pp. 1552--1557.

\bibitem[{Bhatt and Kankanhalli(2011)}]{Bhatt2011}
Bhatt, C.~A., Kankanhalli, M.~S., 2011. Multimedia data mining: State of the
  art and challenges. Multimedia Tools Appl. 51~(1), 35--76.

\bibitem[{Bishop(2006)}]{Bishop2006}
Bishop, C.~M., 2006. Pattern Recognition and Machine Learning (Information
  Science and Statistics). Springer-Verlag New York, Inc., Secaucus, NJ, USA.

\bibitem[{Blank et~al.(2005)Blank, Gorelick, Shechtman, Irani, and
  Basri}]{Blank2005}
Blank, M., Gorelick, L., Shechtman, E., Irani, M., Basri, R., 2005. Actions as
  space-time shapes. In: Computer Vision, 2005. ICCV 2005. Tenth IEEE
  International Conference on. Vol.~2. pp. 1395--1402 Vol. 2.

\bibitem[{Bobick and Davis(1996)}]{Bobick96}
Bobick, A.~F., Davis, J.~W., 1996. An appearance-based representation of
  action. In: Proceedings of the 13th Int. Conf. on Pattern Recognition (ICPR).
  pp. 307--312.

\bibitem[{Bobick and Davis(2001)}]{Bobick01}
Bobick, A.~F., Davis, J.~W., 2001. The recognition of human movement using
  temporal templates. IEEE Transactions on Pattern Analysis and Machine
  Intelligence 23~(3), 257--267.

\bibitem[{Cho et~al.(2009)Cho, Chao, Lin, and Chen}]{Cho09}
Cho, C.~W., Chao, W.~H., Lin, S.~H., Chen, Y.~Y., 2009. A vision-based analysis
  system for gait recognition in patients with parkinson's disease. Expert
  Systems with Applications 36(3)~(3), 7033--7039.

\bibitem[{Díaz-Pereira et~al.(2014)Díaz-Pereira, Gómez-Conde, Escalona, and
  Olivieri}]{Pino2014}
Díaz-Pereira, M.~P., Gómez-Conde, I., Escalona, M., Olivieri, D.~N., 2014.
  Automatic recognition and scoring of olympic rhythmic gymnastic movements.
  Human Movement Science in press.

\bibitem[{Ekinci and Aykut(2007)}]{Ekinci2007}
Ekinci, M., Aykut, M., 2007. Human gait recognition based on kernel pca using
  projections. Journal of Computer Science and Technology 22, 867--876.

\bibitem[{Etemad and Chellappa(1997)}]{Etemad1997}
Etemad, K., Chellappa, R., 1997. Discriminant analysis for recognition of human
  face images. In: Audio and Video-based Biometric Person Authentication. Vol.
  14(8) of Lecture Notes in Computer Science. pp. 1724--1733.

\bibitem[{Farneb\"{a}ck(2003)}]{Farn03}
Farneb\"{a}ck, G., 2003. Two-frame motion estimation based on polynomial
  expansion. In: Proceedings of the 13th Scandinavian Conf. on Image analysis.
  pp. 363--370.

\bibitem[{Felzenszwalb et~al.(2010)Felzenszwalb, Girshick, McAllester, and
  Ramanan}]{Felzenszwalb2010}
Felzenszwalb, P., Girshick, R., McAllester, D., Ramanan, D., 2010. Object
  detection with discriminatively trained part-based models. IEEE Transactions
  on Pattern Analysis and Machine Intelligence 32(9)~(9), 1627 --1645.

\bibitem[{Fukunaga(1990)}]{Fukunaga1990}
Fukunaga, K., 1990. Introduction to statistical pattern recognition (2nd ed.).
  Academic Press Professional, Inc., San Diego, CA, USA.

\bibitem[{Hosseini and Eftekhari-Moghadam(2013)}]{Hosseini2013}
Hosseini, M.-S., Eftekhari-Moghadam, A.-M., 2013. Fuzzy rule-based reasoning
  approach for event detection and annotation of broadcast soccer video.
  Applied Soft Computing 13~(2), 846 -- 866.

\bibitem[{Hu et~al.(2011)Hu, Xie, Li, Zeng, and Maybank}]{Hu2011}
Hu, W., Xie, N., Li, L., Zeng, X., Maybank, S., 2011. A survey on visual
  content-based video indexing and retrieval. Systems, Man, and Cybernetics,
  Part C: Applications and Reviews, IEEE Transactions on 41~(6), 797--819.

\bibitem[{Huang et~al.(1999)Huang, Harris, and Nixon}]{Huang99B}
Huang, P.~S., Harris, C.~J., Nixon, M.~S., aug. 1999. Human gait recognition in
  canonical space using temporal templates. In: IEE Proceedings of Vision,
  Image and Signal Processing. Vol. 146(2). pp. 93--100.

\bibitem[{Ikizler and Forsyth(2008)}]{Ikizler2008}
Ikizler, N., Forsyth, D., 2008. Searching for complex human activities with no
  visual examples. Int. J. Computer Vision 80(3), 337--357.

\bibitem[{Jhuang et~al.(2007)Jhuang, Serre, Wolf, and Poggio}]{Jhuang2007}
Jhuang, H., Serre, T., Wolf, L., Poggio, T., 2007. A biologically inspired
  system for action recognition. In: Computer Vision, 2007. ICCV 2007. IEEE
  11th International Conference on. pp. 1--8.

\bibitem[{Jones and Shao(2013)}]{Jones2013}
Jones, S., Shao, L., 2013. Content-based retrieval of human actions from
  realistic video databases. Information Sciences 236, 56--65.

\bibitem[{Keller et~al.(1985)Keller, Gray, and Givens}]{Keller1985}
Keller, J., Gray, M., Givens, J., July 1985. A fuzzy k-nearest neighbor
  algorithm. Systems, Man and Cybernetics, IEEE Transactions on SMC-15~(4),
  580--585.

\bibitem[{Küçük and Yazıcı(2011)}]{Kucuk2011}
Küçük, D., Yazıcı, A., 2011. Exploiting information extraction techniques
  for automatic semantic video indexing with an application to turkish news
  videos. Knowledge-Based Systems 24~(6), 844 -- 857.

\bibitem[{Lam et~al.(2007)Lam, Lee, and Zhang}]{Lam2007}
Lam, T.~H., Lee, R.~S., Zhang, D., 2007. Human gait recognition by the fusion
  of motion and static spatio-temporal templates. Pattern Recognition 40~(9),
  2563 -- 2573.

\bibitem[{Laptev et~al.(2008)Laptev, Marszalek, Schmid, and
  Rozenfeld}]{Laptev2008}
Laptev, I., Marszalek, M., Schmid, C., Rozenfeld, B., June 2008. Learning
  realistic human actions from movies. In: Computer Vision and Pattern
  Recognition, 2008. CVPR 2008. IEEE Conference on. pp. 1--8.

\bibitem[{Liao et~al.(2013)Liao, Liu, Xiao, and Liu}]{Liao2013}
Liao, K., Liu, G., Xiao, L., Liu, C., 2013. A sample-based hierarchical
  adaptive k-means clustering method for large-scale video retrieval.
  Knowledge-Based Systems 49, 123 -- 133.

\bibitem[{Liu and Shah(2008)}]{Liu2008}
Liu, J., Shah, M., 2008. Learning human actions via information maximization.
  In: Computer Vision and Pattern Recognition, 2008. CVPR 2008. IEEE Conference
  on. pp. 1--8.

\bibitem[{Lucas and Kanade(1981)}]{Lucas1981}
Lucas, B.~D., Kanade, T., 1981. An iterative image registration technique with
  an application to stereo vision. In: Proceedings of the 7th International
  Joint Conference on Artificial Intelligence - Volume 2. IJCAI'81. Morgan
  Kaufmann Publishers Inc., San Francisco, CA, USA, pp. 674--679.

\bibitem[{Luh and Lin(2011)}]{Luh2011}
Luh, G.-C., Lin, C.-Y., 2011. \{PCA\} based immune networks for human face
  recognition. Applied Soft Computing 11~(2), 1743 -- 1752, the Impact of Soft
  Computing for the Progress of Artificial Intelligence.

\bibitem[{Meeds et~al.(2008)Meeds, Ross, Zemel, and Roweis}]{Meeds2008}
Meeds, E., Ross, D., Zemel, R., Roweis, S., 2008. Learning stick-figure models
  using nonparametric bayesian priors over trees. In: Proceedings of the EEE
  Conference on Computer Vision and Pattern Recognition (CVPR). pp. 1--8.

\bibitem[{Mikolajczyk and Uemura(2008)}]{Mikolajczyk2008}
Mikolajczyk, K., Uemura, H., 2008. Action recognition with motion-appearance
  vocabulary forest. In: Computer Vision and Pattern Recognition, 2008. CVPR
  2008. IEEE Conference on. pp. 1--8.

\bibitem[{Mohri et~al.(2012)Mohri, Rostamizadeh, and Talwalkar}]{Mohri2012}
Mohri, M., Rostamizadeh, A., Talwalkar, A., 2012. Foundations of Machine
  Learning. The MIT Press.

\bibitem[{Nga and Yanai(2014)}]{Nga2014}
Nga, D.~H., Yanai, K., 2014. Automatic extraction of relevant video shots of
  specific actions exploiting web data. Computer Vision and Image Understanding
  118, 2--15.

\bibitem[{Olivieri et~al.(2012)Olivieri, G\'{o}mez~Conde, and
  Vila~Sobrino}]{Olivieri2012}
Olivieri, D.~N., G\'{o}mez~Conde, I., Vila~Sobrino, X.~A., 2012.
  Eigenspace-based fall detection and activity recognition from motion
  templates and machine learning. Expert Syst. Appl. 39~(5), 5935--5945.

\bibitem[{Poppe(2010)}]{Poppe10}
Poppe, R., 2010. A survey on vision-based human action recognition. Image \&
  Vision Computing 28~(6), 976--990.

\bibitem[{Ren et~al.(2009)Ren, Singh, Singh, and Zhu}]{Ren2009}
Ren, W., Singh, S., Singh, M., Zhu, Y., 2009. State-of-the-art on
  spatio-temporal information-based video retrieval. Pattern Recognition
  42~(2), 267 -- 282.

\bibitem[{Rius et~al.(2009)Rius, Gonzàlez, Varona, and Roca}]{Rius2009}
Rius, I., Gonzàlez, J., Varona, J., Roca, F.~X., 2009. Action-specific motion
  prior for efficient bayesian 3d human body tracking. Pattern Recognition
  42~(11), 2907 -- 2921.

\bibitem[{Samy~Sadek and Michaelis2(2013)}]{Sadek2013}
Samy~Sadek, Ayoub Al-Hamadi, G.~K., Michaelis2, B., 6 2013. Affine-invariant
  feature extraction for activity recognition. ISRN Machine Vision 2013.

\bibitem[{Schindler and Van~Gool(2008)}]{Schindler2008}
Schindler, K., Van~Gool, L., 2008. Action snippets: How many frames does human
  action recognition require? In: Computer Vision and Pattern Recognition,
  2008. CVPR 2008. IEEE Conference on. pp. 1--8.

\bibitem[{Scholkopf et~al.(1999)Scholkopf, Smola, and Müller}]{Scholkopf1999}
Scholkopf, B., Smola, A., Müller, K.-R., 1999. Kernel principal component
  analysis. In: Advances in kernel methods - support vector learning. MIT
  Press, pp. 327--352.

\bibitem[{Schuldt et~al.(2004)Schuldt, Laptev, and Caputo}]{Schuldt2004}
Schuldt, C., Laptev, I., Caputo, B., 2004. Recognizing human actions: A local
  svm approach. In: Proceedings of the Pattern Recognition, 17th International
  Conference on (ICPR'04). IEEE Computer Society, Washington, DC, USA, pp.
  32--36.

\bibitem[{Szeliski(2010)}]{Szeliski2010}
Szeliski, R., 2010. Computer Vision: Algorithms and Applications, 1st Edition.
  Springer-Verlag New York, Inc., New York, NY, USA.

\bibitem[{Ugolotti et~al.(2013)Ugolotti, Nashed, Mesejo, Špela Ivekovič,
  Mussi, and Cagnoni}]{Ugolotti2013}
Ugolotti, R., Nashed, Y.~S., Mesejo, P., Špela Ivekovič, Mussi, L., Cagnoni,
  S., 2013. Particle swarm optimization and differential evolution for
  model-based object detection. Applied Soft Computing 13~(6), 3092 -- 3105,
  swarm intelligence in image and video processing.

\bibitem[{{Venkatesh Babu} and Ramakrishnan(2004)}]{VenkateshBabu2004}
{Venkatesh Babu}, R., Ramakrishnan, K.~R., 2004. Recognition of human actions
  using motion history information extracted from the compressed video. Image
  and Vision Computing 22(8)~(8), 597--607.

\bibitem[{Xie and Lam(2006)}]{Xie2006}
Xie, X., Lam, K.-M., Sep. 2006. Gabor-based kernel pca with doubly nonlinear
  mapping for face recognition with a single face image. Trans. Img. Proc.
  15~(9), 2481--2492.

\end{thebibliography}

\end{document}